\documentclass{aa}

\usepackage{graphicx}
\usepackage[varg]{txfonts}
\usepackage{natbib,twoopt}
\usepackage[breaklinks=true]{hyperref} 
\bibpunct{(}{)}{;}{a}{}{,} 
\makeatletter
\newcommandtwoopt{\citeads}[3][][]{\href{http://adsabs.harvard.edu/abs/#3}%
{\def\hyper@linkstart##1##2{}%
\let\hyper@linkend\@empty\citealp[#1][#2]{#3}}}
\newcommandtwoopt{\citepads}[3][][]{\href{http://adsabs.harvard.edu/abs/#3}%
{\def\hyper@linkstart##1##2{}%
\let\hyper@linkend\@empty\citep[#1][#2]{#3}}}
\newcommandtwoopt{\citetads}[3][][]{\href{http://adsabs.harvard.edu/abs/#3}%
{\def\hyper@linkstart##1##2{}%
\let\hyper@linkend\@empty\citet[#1][#2]{#3}}}
\newcommandtwoopt{\citeyearads}[3][][]%
{\href{http://adsabs.harvard.edu/abs/#3}
{\def\hyper@linkstart##1##2{}%
\let\hyper@linkend\@empty\citeyear[#1][#2]{#3}}}
\makeatother

\begin{document}

\title{Self-consistent stationary MHD shear flows in the solar atmosphere as electric field generators}

%
\author{D. H. Nickeler\inst{1}, M. Karlick\'y\inst{1},
T. Wiegelmann\inst{2} \and M. Kraus\inst{1}}

\institute{Astronomical Institute, AV \v{C}R, Fri\v{c}ova 298,
25165 Ond\v{r}ejov, Czech Republic\\
\email{dieter.nickeler@asu.cas.cz}
\and
Max-Planck Institut f\"{u}r Sonnensystemforschung, Justus-von-Liebig-Weg 3, 37077 G\"{o}ttingen, Germany
}

\date{Received; accepted}

\authorrunning{Nickeler et al.}
\titlerunning{Stationary MHD shear flows as electric field generators}

\abstract
{Magnetic fields and flows in coronal structures, for example, in gradual phases in
flares, can be described by 2D and 3D magnetohydrostatic (MHS) and steady
magnetohydrodynamic (MHD) equilibria.} 
{Within a physically simplified, but exact mathematical model, we
study the electric currents and corresponding electric fields generated by
shear flows.}
{Starting from exact and analytically
calculated magnetic potential fields, we solveid the nonlinear MHD equations
self-consistently. By applying a magnetic shear flow and assuming a nonideal
MHD environment, we calculated an electric field via Faraday's
law. The formal solution for the electromagnetic field allowed us to compute 
an expression of an effective resistivity similar to the collisionless Speiser
resistivity.} 
{We find that the electric field can be highly spatially structured, or in
other words, 
filamented. The electric field component parallel to the magnetic field
is the dominant component and is high where the resistivity has a maximum. 
The electric field is a potential field, therefore, 
the highest energy gain of the particles can be directly derived from the 
corresponding voltage. In our example of a coronal post-flare scenario 
we obtain electron energies of tens of keV, which are on the same order
of magnitude as found observationally. This energy serves as a source for heating
and acceleration of particles.} 
{}

%

\maketitle

%
\section{Introduction}

Dissipation or acceleration processes of energized particles occur in a variety of 
astrophysical plasma environments. For example, acceleration of charged particles is 
observed in the heliosphere, where anomalous cosmic rays are accelerated to high 
energies \citepads[see, e.g.,][]{2010ApJ...709..963D, 2012SSRv..173..283G}, 
in the Earth magnetotail and aurorae \citepads[e.g.,][]{2012SSRv..173...49B}, and in 
solar flares \citepads[see, e.g.,][]{1998SSRv...86...79M, 2002SSRv..101....1A}
and nanoflares \citepads[e.g.,][]{2011A&A...530A.112B}, where electrons and 
ions are heated and accelerated. These processes
are typically connected with strong electric currents and electric fields. A 
reasonable approach for computing these electric fields and currents is 
provided by the theory of magnetohydrodynamics (MHD). But while MHD simulations
compute solutions only on pre-defined grid points, meaning that values of
the electromagnetic field have to be intrapolated, analytical MHD configurations
have the advantage of providing exact knowledge of the electromagnetic field at 
every point in space. Therefore, exact analytical MHD configurations are ideal as 
background fields in test particle simulations. 

To trigger dissipation, for instance, in the form of Ohmic heating, or 
acceleration, a parallel component of the electric field with respect to the magnetic 
field must exist. Such electric field components parallel to the magnetic field can 
be obtained from nonideal MHD.

The heating of the solar plasma and the acceleration of charged particles during
solar flare events is a long-standing problem. Three different main mechansisms
have been described \citepads[for a detailed review see][]{2002SSRv..101....1A}: 
(i) DC-electric field acceleration, which is
typically connected to magnetic collapse processes \citepads[magnetic reconnection such as 
collapsing magnetic loops and the magnetic mirror effect,][]{2007AdSpR..39.1427K}
or via the Betatron mechanism \citepads{2004A&A...419.1159K}, (ii) stochastic
acceleration caused by wave-particle interaction, so-called weak turbulence
\citepads[e.g.,][]{1998SSRv...86...79M, 1999ApJ...517..700L}, and (iii) 
shock acceleration. 

The scenario of DC-electric field acceleration is very promising, in particular
for the aftermath of a solar flare event, where magnetic reconnection 
had taken place and a plasmoid was ejected. In such a reconnection region, strong 
electric fields are generated, which can directly accelerate charged particles. These
particles are traced by their X-ray emission. However, after the new equilibrium state 
is reached, the main magnetic field component of such a post-flare configuration is
the poloidal magnetic field, which can typically be described by a potential field. This can be
justified by the fact that after the impulsive phase the main component of the field should be
relaxed. However, the \lq bursty\rq~reconnection event itself is not sufficient to explain
the observed slow decay in intensity of the X-ray observations taken 
immediately after the impulsive phase \citepads[see, e.g.,][]{1974IAUS...57..105K}. 
This behavior of the emission speaks in favor of
a continous (although reduced) acceleration on much longer timescales. 
If the relaxed configuration would consist of a pure potential field, this would imply that no further 
dissipation can take place and particle acceleration has stopped, in contrast to what 
is observed. Hence, particles must also be accelerated in the (almost) relaxed magnetic field. 
Therefore, a current-producing shear component must exist,
which provides a reasonably strong electric field component that is necessary to accelerate charged particles along
the field lines. Such shear fields have been observed \citepads[see, 
e.g.,][]{1992SoPh..140...85W}.

In this paper we investigate the influence of shear flows on the generation of 
electric fields with component parallel to the magnetic field in a typical 
post-flare configuration. In numerical test-particle approaches, as has been 
critically commented on by, e.g., \citetads{2009A&A...508..993B} and
\citetads{2011SSRv..159..357Z}, the particles are passive, which means that
the feedback of the moving charges is not taken into account. This could 
be done numerically by considering a kinetic approach. However, because in 
kinetic models spatial and time scales have to be resolved, which requires quite
different scales (Debye length and gyro-frequency), this 
treatment is numerically expensive. In contrast, our exact analytical 
nonideal MHD model allows us to precisely compute the field everywhere, not only
on a predefined grid. In addition, our treatment of the accelerated bulk 
particles automatically includes the nonlinear feedback between the plasma and 
the electromagnetic field, which emphasizes the advantage of exact analytical 
models.


\section{Theoretical approach}

\subsection{Derivation of the MHD model}

In typical simulation scenarios a shear is applied to the footpoints of solar arcade structures 
\citepads[see, e.g.,][and references therein]{2013ApJ...778...99L}. Then the system relaxes into a 
new state. However, this new state is not necessarily an exact equilibrium 
state. Becausewe aim at an exact steady-state for our model considerations, we have to follow a strategy that
allows us to compute the exact final state into which the system relaxes. This 
is offered by the 
transformation theory, which was developed by \citetads{1992PhFlB...4.1689G}. The transformation method allows 
us to calculate steady 
ideal MHD equilibria with field-aligned incompressible flow from known MHS equilibria 
\citepads[see, e.g.,][]{1999GApFD..91..269P, 2006A&A...454..797N, 2013A&A...556A..61N, 
2010AnGeo..28.1523N, 2012AnGeo..30..545N}, and it is applied here to obtain 
a stationary equilibrium, consisting of a poloidal field and a shear component in $z$-direction, from an originally pure potential field.
Our chosen coordinate system is such that the $y$-axis is perpendicular to the 
solar surface (pointing upward), and the $x$-axis is tangential. 
The $z$-axis is tangential as well and points out of the (poloidal) plane in all our 
graphics.

We start from the set of stationary MHD equations for field-aligned, incompressible flows, 
given by
\begin{eqnarray}
  \vec\nabla\cdot\left( \rho \vec v \right) &=& 0\, ,\label{mce}\\
   \rho\left( \vec v\cdot\vec\nabla\right)\vec v &=& \vec
j\times\vec B - \vec\nabla P\, \label{ee},\\
\vec\nabla\times\left(\vec v\times\vec B\right) &=&\vec 0\, ,\label{ie}\\
     \vec\nabla\times\vec B &=& \mu_{0}\vec j\, ,\label{al}\\
      \vec\nabla\cdot\vec B &=& 0\, ,\label{sc}\\
       \vec\nabla\cdot\vec v &=& 0\, ,\label{ice}\\
       \vec v &=& \pm|M_{A}| \vec v_{A}  \\
      \vec v_{A}& := &\frac{\vec B}{\sqrt{\mu_{0}\rho}}\, , \label{letzte}
        \end{eqnarray}
where $\rho$ is the mass density, $\vec{\rm v}$ is the plasma velocity, $\vec B$ is the magnetic flux 
density, $\vec j$ is the current density, $P$ is the plasma pressure, $M_{A}$ is the Alfv\'{e}n Mach 
number, $\vec{\rm v_{A}}$ is the Alfv\'{e}n velocity, and $\mu_{0}$ is the magnetic permeability of 
the vacuum. 
The gravitational force in the considered domains of our model is at least 
a factor 100 lower than the Lorentz-force, so that the influence of
gravity in our approach can be neglected. 

In the current work we focus on solar magnetic arcade structures, implying translational invariance. 
This justifies restricting our investigations to 2.5D magnetic field configurations, that is,
$\partial/\partial z = 0$ for all our variables, although the transformation used could also be applied
to full 3D scenarios.
In addition to the translational invariance, we assume that there is no electric current
component in the invariant (here $z$) direction, that is, $j_{z} = 0$.
As $j_{z} = 0$, we have to solve the Laplace equation $\Delta A = 0$ for the 
flux function $A$ or $\Delta \phi_m = 0$ for the complex conjugated vector 
potential $\phi_m$. This is commonly achieved by complex analysis.
Hence we define by $\phi_{m}$ and 
$A$ the complex conjugated potentials of the complex magnetic vector potential 
\begin{equation}
{\cal A}(u) = \phi_{m} + i A\, ,
\end{equation}
with $u = x + i y$. These potentials fulfill the Cauchy-Riemann equations $\vec\nabla\phi_{m} = \vec 
\nabla A \times \vec e_{z}$ and, therefore, obey the condition $\vec\nabla\phi_{m} \cdot\vec\nabla A = 0$. 
To determine the magnetic potential $\phi_{m}$ and the flux function $A$, the 
Laplace equation is solved by expressing ${\cal A}(u)$ with Laurent series, 
applying asymptotical boundary conditions.

With the magnetic potentials, we can use the potential field $\vec\nabla\phi_{m}$ to define  
the static poloidal magnetic field $\vec B_{ps}$ in the form
\begin{equation}
\vec B_{ps} = \sqrt{1-M_{A}^{2}}\vec\nabla\phi_{m} = \sqrt{1-M_{A}^{2}}\vec B_{p}\, ,
\end{equation}
where $\vec\nabla\phi_{m} = \vec B_{p}$ is the stationary poloidal magnetic field and 
$M_{A}$ is the constant Alfv\'{e}n Mach number. The requirement of $M_{A}=\textrm{const}$ follows from the fact
that on the one hand, $\vec B_{ps}$ is a potential field, and on the other hand, no current is generated in 
$z$-direction by the transformation. For simplicity, to have the representation of the stationary magnetic field
in the usual form via $\vec\nabla\phi_{m}$, we introduced the factor $\sqrt{1-M_{A}^{2}}$ into the static 
poloidal magnetic field, which vanishes identically after the transformation
has been applied. For the current investigation, we considered only 
sub-Alfv\'{e}nic flows, which means that $M_{A}^{2} < 1$. The initial potential magnetic field 
$\vec B_{ps}$ together with the static plasma pressure $p_{s0} = \textrm{const}$ 
define the starting MHS equilibrium.

To compute the stationary MHD equilibrium, we applied a 2.5D shear flow $\vec v$ and simultaneously 
performed the transformation, so that we obtained a self-consistent 2.5D MHD flow. While the $x$ and $y$ 
components of the shear flow are functions of the poloidal magnetic field, the $z$ component produces 
a nonconstant magnetic shear $B_{z} = B_{z}(A)$ in $z$-direction, given by 
\begin{equation}
B_{z}(A) = \frac{v_{z}(A)}{M_{A}}~\sqrt{\mu_{0}\rho(A)}\, , \label{shearflow}
\end{equation}
where $\rho(A)$ is the plasma density. As was shown by \citetads{2006A&A...454..797N}, the density is an explicit
function of $A$. To fulfill the requirement of a field-aligned flow, the shear flow has to have
the following structure
\begin{equation}
\vec v = \frac{M_{A}}{\sqrt{\mu_{0}\rho(A) (1-M_{A}^{2})}} \vec B_{ps} + \frac{M_{A}}{\sqrt{\mu_{0} \rho(A)}}
B_{z}(A)\vec e_{z}\, .\label{flow}
\end{equation}
From the transformation and the application of the shear the stationary magnetic field has the form
\begin{equation}
\vec B = \vec B_{p} + B_{z}(A) \vec e_{z} \equiv \vec\nabla\phi_{m} + B_{z}(A) \vec e_{z}\, ,\label{magfield}
\end{equation}
and the electric current is given by Amp\`{e}re's law, 
\begin{eqnarray}
\vec j & = & \frac{1}{\mu_{0}}\vec\nabla\times B_{z}\vec e_{z} 
             =  \frac{1}{\mu_{0}}\vec\nabla B_{z}\times \vec e_{z} \\
       & = & \frac{1}{\mu_{0}} B_{z}'(A)\vec\nabla A \times \vec e_{z} 
             = \frac{1}{\mu_{0}} B_{z}'(A)\vec\nabla \phi_{m}\, .
\label{currentidentity}
\end{eqnarray}
The prime denotes the derivative with respect to $A$, that is, $B_{z}'(A) = \mathrm{d}B_{z}/\mathrm{d}A$.
The thermal pressure of the sheared equilibrium is
\begin{equation}
p=p_{s} - \frac{\rho(A)}{2} \, \vec{v}\,^{2} = p_{s0}-\frac{1}{2\mu_{0}}\, B_{z}^2(A)-\frac{M_{A}^2}{2\mu_{0}}\,|\vec\nabla A|^2\, . \label{druck}
\end{equation}
The parameter $p_{s}$ in Eq.\,(\ref{druck}) is the static pressure of an equilibrium state like ours plus a shear
component. This means that we define the total static pressure as $p_{s} = p_{s0} - (1-M_{A}^{2}) B_{z}^{2}/(2 \mu_{0})$,
with $p_{s0}$ as the static pressure of the pure poloidal field. 
The thermal pressure (Eq.\,(\ref{druck})) together with the flow (Eq.\,(\ref{flow})) and the magnetic field 
(Eq.\,(\ref{magfield})) self-consistently fulfill the incompressible {\it ideal} steady-state MHD equations with field-aligned flows.

In ideal MHD, Ohm's law is given by
\begin{equation}
\vec E + \vec v \times \vec B = \vec 0\,. 
\end{equation}
The use of field-aligned flows, which implies that $\vec v \times \vec B = \vec 0$, has a severe
consequence, because such an electro-magnetic field configuration in ideal MHD cannot
accelerate injected charged particles because the electric field is zero. 
To guarantee particle acceleration, we must rely on resistive MHD, using resistive Ohm's law given by
\begin{equation}
\vec E + \vec v \times \vec B = \eta \vec j\, .
\end{equation}
In this scenario an electric field also exists for field-aligned flows, but only for $\eta \vec 
j \neq 0$ and hence in particular for $\eta \neq 0$.

To enable quasi-steady and sustainable electric fields with field-aligned flows,
it is necessary to simultaneously solve the resistive Ohm's law and the momentum
equation. This was first investigated by \citeads{1970PhRvL..24.1337G} and 
subsequently by \citetads{1998JPlPh..59..303T} and \citetads{2000JPlPh..64..601T} for the
case of axisymmetric fields. These authors found several classes of analytical equilibria 
with physically plausible $\sigma = 1/\eta$ profiles, stating clearly that in general, 
fields with constant resistivity do not exist by proving that only the assumption 
of a spatially varying $\sigma$ makes the equilibrium problem well posed. In addition, 
they found that according to Ohm's law $\eta$ is determining the MHD solutions, but $\eta$ is also 
determined by constraints concerning the geometry of flow and field. Therefore, 
we applied a similar concept of a non-constant resistivity for our
translational invariant model.

The steady-state Maxwell equation $\vec\nabla\times\vec E = -\partial \vec B/\partial t = \vec 0$ implies
the existence of an electric potential $\phi_{e}$ so that $-\vec\nabla\phi_{e} = \vec E = \eta\vec j$. 
Because the Cauchy-Rieman differential equations imply that $\vec\nabla\phi_{m}\cdot\vec\nabla A = 0$, we can choose 
$\phi_{m}$ and $A$ as orthogonal coordinates. This allows us to consider  
the electric potential $\phi_{e}$ as a function of the magnetic potential field $\phi_{m}$ and 
the flux function $A$. With the current, given by Eq.\,(\ref{currentidentity}), we can re-write
the identity $\eta\vec j = -\vec\nabla\phi_{e}$ in the form
\begin{equation}
\frac{1}{\mu_{0}}\eta(\phi_{m},A) B_{z}'(A)\vec\nabla\phi_{m} = -\frac{\partial \phi_{e}}{\partial
\phi_{m}} \vec\nabla\phi_{m} - \frac{\partial \phi_{e}}{\partial A}\vec\nabla A\, .
\end{equation}
A comparison of the coefficients delivers that $\partial \phi_{e}/\partial A = 0$, which means that 
the electric potential $\phi_{e}$ is an explicit function of $\phi_{m}$ only, and therefore
\begin{equation}
\frac{1}{\mu_{0}}\eta(\phi_{m},A) B_{z}'(A) = - \frac{\mathrm{d}\phi_{e}(\phi_{m})}{\mathrm{d}\phi_{m}} =: 
\frac{1}{\mu_{0}}\xi(\phi_{m})\, ,
\label{elecpot}
\end{equation}
where $\xi(\phi_{m}) = \eta(\phi_{m},A) B_{z}'(A)$, which has the SI unit Ohm, can be regarded as the resistance of the plasma.
Furthermore, we find from Eq.\,(\ref{elecpot}) for the electric potential $\phi_{e}$
\begin{equation}
\phi_{e}=-\frac{1}{\mu_{0}\,}
\displaystyle\int\xi(\phi_{m})\, d\phi_{m}\, .
\label{otherelecpot}
\end{equation}
Consequently, equipotential surfaces of the scalar potential $\phi_{m}$ of the poloidal 
magnetic field are also equi\-potential surfaces of the electric potential, 
$\phi_{e}=\phi_{e}(\phi_{m})$.

Considering the different representations, the electric field can be written in various equivalent forms:
\begin{eqnarray}
\vec E & = & \eta\vec j=\frac{\eta}{\mu_{0}}\, B_{z}'(A)\,\vec\nabla \phi_{m} =
\frac{1}{\mu_{0}}\,\xi(\phi_{m})\,\vec\nabla\phi_{m} \label{Evarious1}\\
  & = & \frac{1}{\mu_{0}}\,
\vec\nabla\displaystyle\int\xi(\phi_{m})\, d\phi_{m}\, = - \vec\nabla \phi_{e}. \label{Evarious}
\end{eqnarray}
The existence of an electric field component aligned with the magnetic field provides an energy gain and 
hence an acceleration of charged particles along the field lines. The computation of the total 
electric field (and hence the parallel component) explicitly depends on the resistivity $\eta$, but implicitly 
on the resistance $\xi$. 
In the scenario of a post-flare configuration described above, which neglects currents in the 
invariant direction and an electric field component produced by the flow, we found that the resistivity is 
given by (see Eq.\,(\ref{elecpot}))
\begin{equation}
\eta(\phi_{m},A) = \frac{\xi(\phi_{m})}{B_{z}'(A)}\, \label{etadef} .
\end{equation}
Although the resistivity is an explicit function of the two complex conjugated potentials $\phi_{m}$ and $A$,
it is very special in the sense that the two coordinates appear separately in the two functions $\xi$ and 
$B_{z}'$ that determine $\eta$.

The resistance, $\xi$, and the derivative of the magnetic shear, $B_{z}'$, can basically be chosen 
independently. To investigate the properties of this resistivity, we discuss 
various options for functions $\xi$ and $B_{z}'$ in the following.
The case of a constant magnetic shear component can directly be excluded 
because this would imply that $B_{z}'(A)=0$ and hence the configuration would contain no currents and the resistivity would be undefined. For a nonconstant magnetic shear component we are left with four different possibilities: (i) $B_{z}'(A)$ is constant and $\xi$ is constant; (ii) $B_{z}'(A)$ is constant and $\xi$ is not constant;
(iii) $B_{z}'(A)$ is nonconstant but $\xi$ is constant; and (iv) $B_{z}'(A)$ and $\xi$ are both nonconstant. 

If $\xi$ and $B_{z}'(A)$ were both constant, the resistivity would be constant as well, meaning that the electric field 
only depends on the poloidal magnetic field (see Eq.\,(\ref{Evarious1})). 
Hence, a strong electric field would only occur in regions of high magnetic field 
strength, for example, in regions of high current density (see Eq.\,(\ref{currentidentity})). Such regions of high current density occur close to the poles of 
multipolar regions, for instance. If $B_{z}'(A)$ is constant and $\xi$ is not constant, $\eta$ only depends on $\phi_{m}$ and only varies in 
direction along the magnetic field lines. 
If $B_{z}'(A)\neq$~constant, $\eta$ will decrease with increasing $B_{z}'(A)$,
but the choice of the function $\xi$ enables us to receive a high
resistivity at specific locations.
Hence choosing a constant $\xi$ is unsuitable because it prevents a change
and localization of the electric field along the magnetic field lines and therefore 
an effective concentration of the particle acceleration engine.

The search for appropriate expressions for the resistivity is additionally hampered by the
fact that in cases of nonconstant $\xi$ one has to find a physical explanation for $\xi$, which
is not obvious a priori. Evidently, $\xi$ represents a characteristic electric resistance, which 
depends on the distance along the magnetic field lines.

A high resistivity alone is not sufficient for the occurrence of strong (parallel) electric fields. 
Instead, appropriate choices for the spatial distributions of both the poloidal magnetic field 
and the function $\xi$ are required. The regions where the conditions for the appearance of strong electric 
fields are met are, therefore, not necessarily identical to those where the effective resistivity 
$\eta$ is particularly high. Nevertheless, we need a physical explanation for the resistivity and
a reasonable physical correlation between resistivity and the function $\xi$. The Spitzer resistivity
is not valid for solar corona or solar flare scenarios because the density is 
too low, which prevents efficient 
collisions between particles. On the other hand, the turbulent collisionless resistivity 
(anomalous resistivity due to wave-particle interactions), which occurs during eruptive procsses such as 
impulsive phases of flares, is usually not stationary. Therefore, the most appropriate approach for 
our investigations is to find a stationary resistivity that take noncollisional effects into account.

\subsection{Geometrical approach for the noncollisional inertia-induced resistivity: Speiser-like approach}

The resistivity in our model depends on the resistance 
$\xi$, which was defined in Eq.\,(\ref{elecpot}) as a function of the vector potential $\phi_{m}$. 
The resistance $\xi(\phi_{m})$ and the resistivity $\eta=\xi(\phi_{m})/B'_{z}(A)$ define formally exact solutions for the resistivity and the electric field
in the specific geometry of a given magnetic field, but for estimating the \lq amplitude\rq~, that is,
the absolute value of $\xi$ and thus $\eta$, a quantitative approach is required.
To estimate a noncollisional but steady-state $\eta$ it is hence essential to derive a parameterization 
for the resistance. For this, we make a short excursion into the two-fluid approach. Here, 
the resistance $\xi$ can be represented by utilizing the ideas 
of the so-called Speiser conductivity \citepads{1970P&SS...18..613S}, which is based on the inertia of the injected 
and accelerated charged particles. This means that we are using stationary movements of particles to 
derive the relation between the total electric fields (see Eqs.\,(\ref{Etot1}) and (\ref{Etot2}) 
below) and $\xi$. This procedure can be justified by the suggestion of Speiser 
\citepads[see, e.g.,][]{1968epf..conf..393S, 1970P&SS...18..613S, 1969P&SS...17.1285D},
who stated that the effect of particle inertia can be more important 
in determining the relation between $\vec E$ and $\vec j$ in a collisionless plasma than wave-particle 
interactions.

Differently from the situation in the Speiser models, the geometry used
in our models is more complicated: our models are 2.5D while
the Speiser models were 1.5D (see e.g. Lyons \& Speiser, 1985).
In addition, our current sheet is not given by the Maxwellian jump condition
with respect to the {\sl main component} $B_{p}$ of the magnetic field, but by
the {\sl shear component} $B_{z}$. 

In our model we assume that a steady-state flow of coronal material can develop in which 
the plasma streams along the field lines that have been sheared in $z$-direction. However, within 
this ordered plasma stream, a drift between the different species of charged particles 
(electrons and protons or ions) can be expected. The drift inside the global plasma flow initiates 
a current, which is, via Amp\`{e}re's law, related to a magnetic shear component. Hence, no 
turbulent approach is needed to obtain an electric field. Instead, it results naturally from the 
inertial forces.

The inertial forces acting on the charged particles generated by the electric and magnetic
fields can be written as
 \begin{eqnarray}
\vec E+ \vec v_{i}\times\vec B &=&\frac{m_{i}\vec a_{i}}{q}\, 
 :=\vec E_{i}\, , \label{Etot1} \\ 
\vec E+ \vec v_{e}\times\vec B &=&-\frac{m_{e}\vec a_{e}}{e}\, 
  :=\vec E_{e}\, , \label{Etot2}
\end{eqnarray}
where $\vec v_{i}$ and $\vec v_{e}$ are the velocity fields of the charged particles (ions $i$ and electrons $e$),
$m_{i}$ and $m_{e}$ are their masses, $\vec a_{i}$ and $\vec a_{e}$ are the 
accelerations acting on the bulk of particles, and $q$ and $e$ are their charges.
The coupling between the Maxwell equation (Amp\`{e}re's law) and the fluid equations
is realized via Eqs.\,(\ref{drift}, \ref{stromshear}) and (\ref{coupling}) below.
Here, we explicitly considered only the electric force and the Lorentz force, while
all other collective forces (like $\vec\nabla P_{e}$, the Hall term, etc.) are included in the
total electric field for each species (ions $\vec E_{i}$ and electrons $\vec E_{e}$).
The forces $\vec F_{i,e} = m_{i,e}\vec a_{i, e}$ are proportional to the total
electric fields $E_{i,e}$ (= the electric field that particles encounter in the
comoving frame). If we solve for the velocity fields $\vec v_{i,e}$, the 
general solution of the force equations (\ref{Etot1}) and (\ref{Etot2}) is
\begin{eqnarray}
\vec v_{i} & = & \frac{\left(\vec E-\vec E_{i} \right)\times\vec
      B}{|\vec B|^2}+\lambda_{i}\vec B\label{iogeschw}\, ,\\ 
\vec v_{e} & = & \frac{\left(\vec E-\vec E_{e}\right)\times\vec
      B}{|\vec B|^2}+\lambda_{e}\vec B\, .\label{elgeschw}
\end{eqnarray}
The free parameters $\lambda_{i, e}$ of the general solution correspond to
the Alfv\'en Mach number of the flows of the particular particle species:
\begin{eqnarray}
  M_{i}^2:=\lambda_{i}^2\, \mu_{0} n_{i}q\, ,\qquad \textrm{and}\qquad
  M_{e}^2:=\lambda_{e}^2\, \mu_{0} n_{e}e\, .
\end{eqnarray}
The electric current is typically given by the drift between the different charges
\begin{equation}
 \vec j:=n_{i} q\vec v_{i}-n_{e} e\vec v_{e}\, . \label{drift}
\end{equation}
On the other hand, in the MHD picture the current is generated by the magnetic shear 
(see Eq.\,(\ref{currentidentity}))
\begin{equation}
 \vec j = \frac{1}{\mu_{0}}\,B_{z}'(A)\vec\nabla\phi_{m}\, . \label{stromshear}
\end{equation}
As both expressions have to be equal, we can compare the coefficients. For 
this, we introduce a new orthogonal coordinate system defined by the basis $\left( (
\vec\nabla A)^0, \vec e_{z}, (\vec\nabla\phi_{m})^0\right)$. The 
superscript $0$ denotes that the vectors are normalized. Then we can express the physical 
parameters in this new coordinate system: The poloidal magnetic field can be written as 
$\vec B_{p} = B_{p}(\vec\nabla\phi_{m})^{0}$, and the total electric fields that are needed to 
compute the particle velocities become 
\begin{equation}
   \vec E_{i,e}=\alpha_{i,e} (\vec\nabla A)^0+\beta_{i,e}\vec
    e_{z}+\gamma_{i,e} (\vec\nabla\phi_{m})^0\, ,
\end{equation}
where $\alpha, \beta$, and $\gamma$ are the coordinates of the corresponding basis.
The relation for the current density can thus be written in the form
\begin{equation}
 \vec j:=n_{i} q\vec v_{i}-n_{e} e\vec v_{e}\stackrel{!}{=}\frac{1}{\mu_{0}}\,B_{z}'\, B_{p}
   (\vec\nabla\phi_{m})^0\, . \label{coupling}
\end{equation}
The expressions for the velocities, Eqs.\,(\ref{iogeschw}) and (\ref{elgeschw}),
and a comparison of coefficients with respect to the orthogonal unit vectors
result in three conditions for the electric field components of the two
species:
\begin{eqnarray}
  j_{A} &=& \frac{n_{e}e - n_{i}q}{\mu_{0}|\vec B|^2}\, \xi B_{z}
    B_{p} - \frac{n_{i}q\beta_{i} B_{p}}{|\vec B|^2} + \frac{n_{i}q\gamma_{i}
    B_{z}}{|\vec B|^2} \nonumber\\
   & & + \frac{n_{e} e\beta_{e} B_{p}}{|\vec B|^2} -
    \frac{n_{e} e \gamma_{e} B_{z}}{|\vec B|^2}\,, \label{stromfluko1}\\
  j_{z} &=& \frac{n_{i}q\alpha_{i} B_{p}}{|\vec B|^2} - \frac{n_{e}q
    \alpha_{e} B_{p}}{|\vec B|^2} + \left( n_{i}q\lambda_{i}-n_{e}e\lambda_{e}
    \right) B_{z}\, ,\\
  j_{\phi_{m}} &=&  -\frac{n_{i}q\alpha_{i} B_{z}}{|\vec B|^2} +
    \frac{n_{e} e\alpha_{e} B_{z}}{|\vec B|^2} + \left( n_{i}q\lambda_{i}-
    n_{e}e\lambda_{e}  \right) B_{p}\, ,
\label{stromfluko3}
\end{eqnarray}
with the projection in the direction of the three coordinates of the three basis vectors, namely
\begin{eqnarray}
  j_{A} & :=  & \vec j\cdot (\vec\nabla A)^0 = 0\, , \\
  j_{z} & :=  & \vec j\cdot\vec e_{z} = 0\, , \\
  j_{\phi_{m}} & := & \vec j\cdot (\vec\nabla\phi_{m})^0
    = \frac{1}{\mu_{0}} B_{z}'B_{p}\, .
\end{eqnarray}
Because only $j_{A}$ contains the resistance $\xi$, we concentrate on this
component to receive a relation coupling the physical parameters 
$B_{p}$, $B_{z}$, $\xi$, and $\eta$. This coupling can only be 
calculated if $n_{e}e - n_{i}q$ in the nominator of the first term on the 
right-hand side of Eq.\,(\ref{stromfluko1}) does not vanish. This condition
is only fullfilled as long as $n_{i}\neq n_{e}$ with $n_{i}\approx n_{e}$.

We define the following conductivities:
\begin{equation}
\sigma_{g,e}=\frac{n_{e} e}{|\vec B|}\, \qquad \textrm{and} \qquad \sigma_{g,i}=\frac{n_{i} q}{|\vec B|}\, , \label{def_gyros}
\end{equation}
which resemble the gyroconductivity introduced by \citetads{1968epf..conf..393S}. However, in the definition
of Speiser's gyroconductivity only the $B_{z}$ component was used. This component was oriented
perpendicular to the antiparallel field lines of the magnetic neutral sheet, so that particles within
the neutral sheet are gyrating around $B_{z}$ and therefore encounter an
electric field \citepads[in their own frame of reference, see][]{1985JGR....90.8543L}. 
Since, in our case, there exists no magnetic neutral sheet, we have to
use $|\vec B| = \sqrt{B_{p}^2+B_{z}^2}$ for the corresponding magnetic field, 
and the particles gyrate around its field lines.

By inserting the definition of the gyroconductivities (Eq.\,(\ref{def_gyros})) into the
equation of the current (Eq.\,(\ref{stromfluko1})), we obtain 
\begin{equation}
0=\left(\sigma_{g,e}- \sigma_{g,i} \right) \frac{\xi
 B_{p}}{\mu_{0}}\,\frac{B_{z}}{|\vec B|} 
 + \frac{\sigma_{g,i}}{|\vec B|}( B_{z} \gamma_{i} - B_{p} \beta_{i})
 + \frac{\sigma_{g,e}}{|\vec B|}( B_{p} \beta_{i} - B_{z} \gamma_{i})\, ,\label{gyrolei1}
\end{equation}
from which $\xi$ can in principle be calculated. 
In the situation in which $\beta_{i}=\beta_{e}\neq \gamma_{i}=\gamma_{e}$,
the resistance $\xi$ results in
\begin{equation}
\xi=\mu_{0}\,\frac{B_{z}\gamma_{i} - B_{p}\beta_{i}}{B_{p} B_{z}}\, .
\end{equation}
This scenario is very unlikely, as it implies that the electrons (owing to their 
much smaller mass) would run away from the ions because the electric field accelerates 
the lighter electrons much more strongly than the heavy ions. Hence, the configuration would
not reach a stationary state.

For the ions to achieve about the same acceleration as 
the electrons, the total electric field encountered by the electrons must be negligible 
compared with the total field encountered by the ions.
Therefore, if we set in Eq.\,(\ref{gyrolei1}) $\beta_{i}\gg\beta_{e}$ and 
$\gamma_{i}\gg\gamma_{e}$, which is physically more reasonable, we obtain the 
following expression for $\xi$\, :
\begin{eqnarray}
\xi= \mu_{0}\, \frac{\sigma_{g,i}}{\sigma_{g,e}-\sigma_{g,i}}\,\frac{B_{z}
\gamma_{i}-B_{p}\beta_{i}}{B_{z} B_{p}}\, .
\end{eqnarray}
With the help of this formula we can estimate the
Speiser-like resistivity. 

To ensure that the poloidal component of $\vec E_{i}$ is the dominating one, we demand that
\begin{equation}
\gamma_{i}\approx |\vec E_{i}| = \eta_{i} |\vec j| =  \frac{\eta_{i}}{\mu_{0}} \frac{\mathrm{d}B_{z}(A)}{\mathrm{d}A}
B_{p}\, ,\label{gammai}
\end{equation}
where the term on the right-hand side again refers to the current of the MHD picture.
Then, Eq.\,(\ref{gammai}) can be rewritten in the form
\begin{equation}
\frac{\gamma_{i}}{B_{p}} = \frac{1}{\mu_{0}
  \sigma_{g,i}} \frac{\mathrm{d}B_{z}(A)}{\mathrm{d}A}\, ,
\end{equation}
where we used the identity for the resistivity $\eta_{i} \equiv \frac{1}{\sigma_{g,i}}$.
Because we have no precise information about the acceleration terms from MHD theory, 
it is physically reasonable to request that the acceleration should be field-aligned, that is,
\begin{equation}
\frac{\gamma_{i}}{B_{p}}\approx \frac{\beta_{i}}{B_{z}}\, .
\end{equation}
We now define
\begin{equation}
\frac{\beta_{i} B_{p}}{\gamma_{i} B_{z}} := \varepsilon_{1} \approx 1\, . \label{eps1}
\end{equation}
Therefore, $\xi$ can be written as
\begin{eqnarray}
\xi & = & \mu_{0}\, \frac{\sigma_{g,i}}{\sigma_{g,e}(1-\frac{\sigma_{g,i}}
{\sigma_{g,e}})}\,\frac{\gamma_{i}}{B_{p}} (1 - \varepsilon_{1}) \\
    & = & \mu_{0}\, \frac{\sigma_{g,i}}{\sigma_{g,e}(1-\varepsilon_{2})}
\frac{\gamma_{i}}{B_{p}} (1 - \varepsilon_{1}) \\
    & = & \mu_{0}\, \frac{\sigma_{g,i}}{\sigma_{g,e}(1-\varepsilon_{2})}
\frac{B_{z}'(A)}{\mu_{0}\sigma_{g,i}} (1 - \varepsilon_{1}) 
     =  \frac{1-\varepsilon_{1}}{1-\varepsilon_{2}}
\frac{B_{z}'(A)}{\sigma_{g,e}}\, \label{xi_phys},
\end{eqnarray}
where we introduced the parameter $\varepsilon_{2} = \sigma_{g,i}/\sigma_{g,e}$.
Because according to Eq.\,(\ref{etadef}) $\xi = \eta B_{z}'(A)$, the resistivity has the form
\begin{equation}
\eta = \frac{1-\varepsilon_{1}}{1-\varepsilon_{2}}\,\frac{1}{\sigma_{g,e}} =: 
\varepsilon \,\frac{1}{\sigma_{g,e}}\, .
\label{etaspeiser}
\end{equation}
Owing to the quasi-neutrality $\varepsilon_{2}\approx 1$, which means that
$\eta \sim 1/\sigma_{g,e} = \eta_{e}$. Consequently, the total resistivity, and therefore
also the function $\xi$, is related to the gyroresistivity of the electrons. 

The resistivity (hence the Ohmic heating) depends on the chosen geometry 
and on the constraint of only having almost field-aligned forces.
To keep the resistivity finite, a deviation from exact neutrality ($=$quasi-neutrality) is
required. To keep this deviation small, the field-aligned forces have to be chosen 
appropriately, so that in the expression of the
resistivity, Eq.\,({\ref{etaspeiser}}), the \lq geometrical\rq~term
$(1-\varepsilon_{1})$, which describes the deviation from field-aligned 
acceleration,
compensates for the denominator $(1-\varepsilon_{2})$,
which describes the deviation from perfect neutrality.
This compensation should be made in such a way that
$\eta$ is bounded, implying that
$\varepsilon$ is on the order of unity or at least bounded by
some finite value. This guarantees that the resistivity $\eta$ can be computed at all.
A second important criterion why $\varepsilon$ should be bounded and on the order of unity 
comes from calculating the energies of the charged particles.
Because according to Eq.\,(\ref{otherelecpot}) the electric potential results from 
integrating over the resistance $\xi$,
the magnitude of the voltage that the electric particles can achieve is largely determined
by the value of $\varepsilon$. If $\varepsilon$ were be much higher than unity, the
energies of the particles would approach the relativistic regime, in which the 
theory is not valid anymore.

In the following section we present one example with representative
physical parameters and simplified magnetic field configuration.

\section{Results}

 \begin{figure} 
 \centering
\includegraphics[width=\hsize]{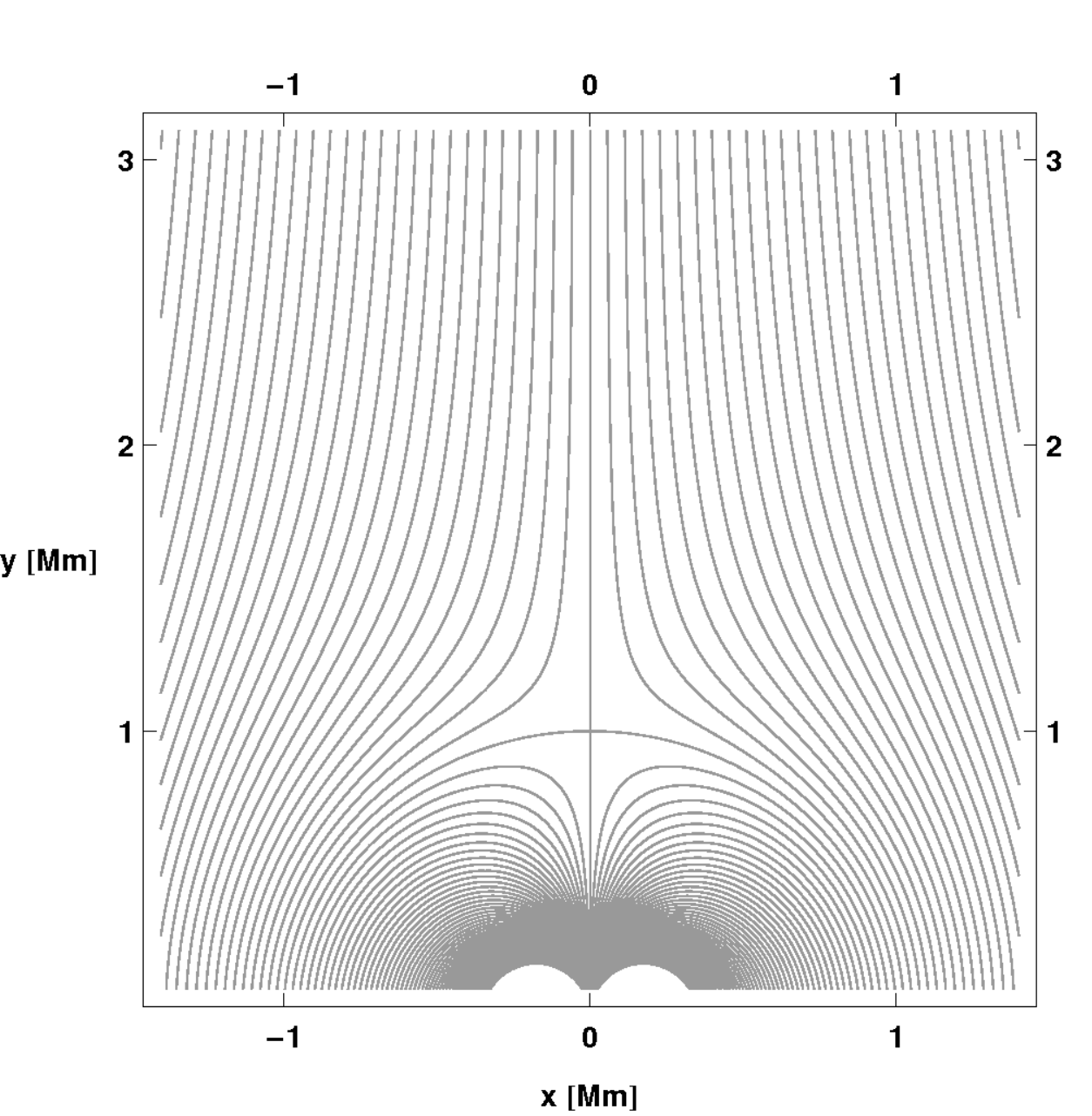}
 \caption{Field line projection onto the poloidal plane.}\label{fig:fili}
 \end{figure}

\begin{figure}
\centering
\includegraphics[width=\hsize]{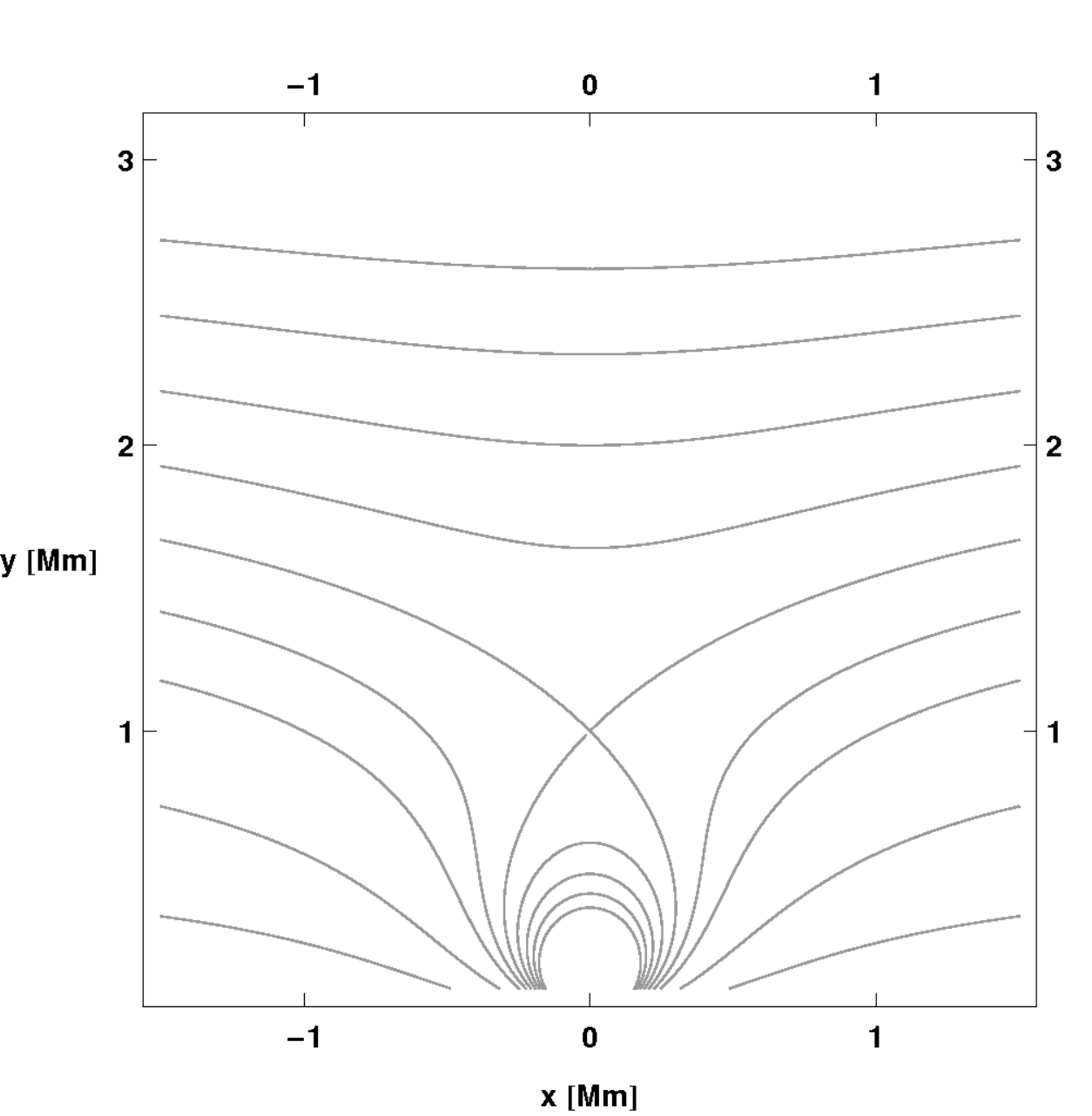}
 \caption{Contour lines of the magnetic potential.}\label{fig:contourphi}
 \end{figure}

The scenario under investigation assumes that the field is almost relaxed,
which supports the assumption of a potential field for the poloidal component of the field (here: the $x$ and $y$ 
components in the translational invariant magnetic field configuration). We concentrate on a small 
region around the footpoint of a solar arcade structure with
dimensions 3\,Mm $\times 3$\,Mm. This domain was chosen to avoid bipolar magnetic field structures with high plasma $\beta$ that contains current sheets with a component of the current in $z$-direction.
We represent the magnetic field configuration with a 2D dipole superimposed on a homogeneous, monopolar field 
component in $y$-direction. This results in a dome-like structure at the bottom and an X-point separatrix on top. 
This configuration is achieved by expanding
the complex magnetic vector potential, ${\cal A}$, in a Laurent series of the form
\begin{equation}
{\cal A} = \sum\limits_{k=-\infty}^{k=\infty} C_{k}u^{k} = -i B_{\infty} u + \frac{C}{u}\, ,
\end{equation}
with the complex constants $C_{k}$ of which only those for the homogeneous component ($C_{1} = -i B_{\infty}$) 
and the dipole component ($C_{-1} = i |C|$) are considered, while all others are set to zero. 
The constant $|C|$ in the latter is given by $|C| = 
B_{\infty} R^{2}$, where $R$ corresponds to the height $y$ above the dipole at which the poloidal field 
vanishes. This height marks the location of the magnetic null point. The choice of the constants guarantees that the asymptotical boundary condition, namely $\vec B(|u|\rightarrow\infty) = B_{\infty} \vec e_{y}$, is fulfilled.
The field lines of the configuration are
displayed in Fig.\,\ref{fig:fili}. Because this figure shows the projection of the field lines onto the poloidal 
plane, it represents the two cases before and after the magnetic shear is applied that results from the shear flow 
(see Eq.\,(\ref{shearflow})), is applied. To compute these field lines,
we chose values appropriate for the solar corona. Magnetic field determinations are usually complicated 
and values obtained from observations at different locations and wavelength regions range from
about $10^{-3}$\,T to $10^{-1}$\,T \citepads[see, e.g.,][]{2000ApJ...541L..83L, 
2006ApJ...641L..69B, 2010ApJ...725L.161C}. We adopted a mean value of the magnetic field of $B_{\infty} = 10^{-2}$\,T
and for the height $R = 1$\,Mm. The contour lines of the magnetic potential $\phi_{m}$ are depicted in 
Fig.\,\ref{fig:contourphi}. They are everywhere perpendicular to the magnetic field lines.

The only effective electric field component that can act
as a particle accelerator is the component parallel to the magnetic field.
From the representations of the total magnetic field (Eq.\,(\ref{magfield})) and the electric field
(Eq.\,(\ref{Evarious})), the parallel component of the electric field, $E_{\parallel}$ , results in
\begin{equation}
    E_{\parallel} = \frac{\vec E\cdot \vec B}{|\vec B|} =
    \frac{1}{\mu_{0}}\,\xi(\phi_{m})\frac{|\vec\nabla\phi_{m}|}{
    \sqrt{1+B_{z}^{2}(A)/|\vec\nabla\phi_{m}|^{2}}}\, .
\end{equation}
To maximize $E_{\parallel}$, it is essential to choose a rather small magnetic shear 
component\footnote{The $B_{z}$ component is not really restricted to low 
values as every $B_{z} = B_{z}(A)$    
produces an equilibrium because of the noncanonical transformation method.
But low values of the shear component compared with the main
poloidal component of the field are of advantage to maximize $E_{\parallel}$,
to justify the assumption of quasi-neutrality, and to fulfill the dominance of the poloidal over the 
$z$-component of the total electric field of the ions (see Eq.\,(\ref{eps1})).}. 
However, the magnetic shear should neither be constant nor zero, because this
would imply a current-free field.

\begin{figure}
\centering
 \includegraphics[width=\hsize]{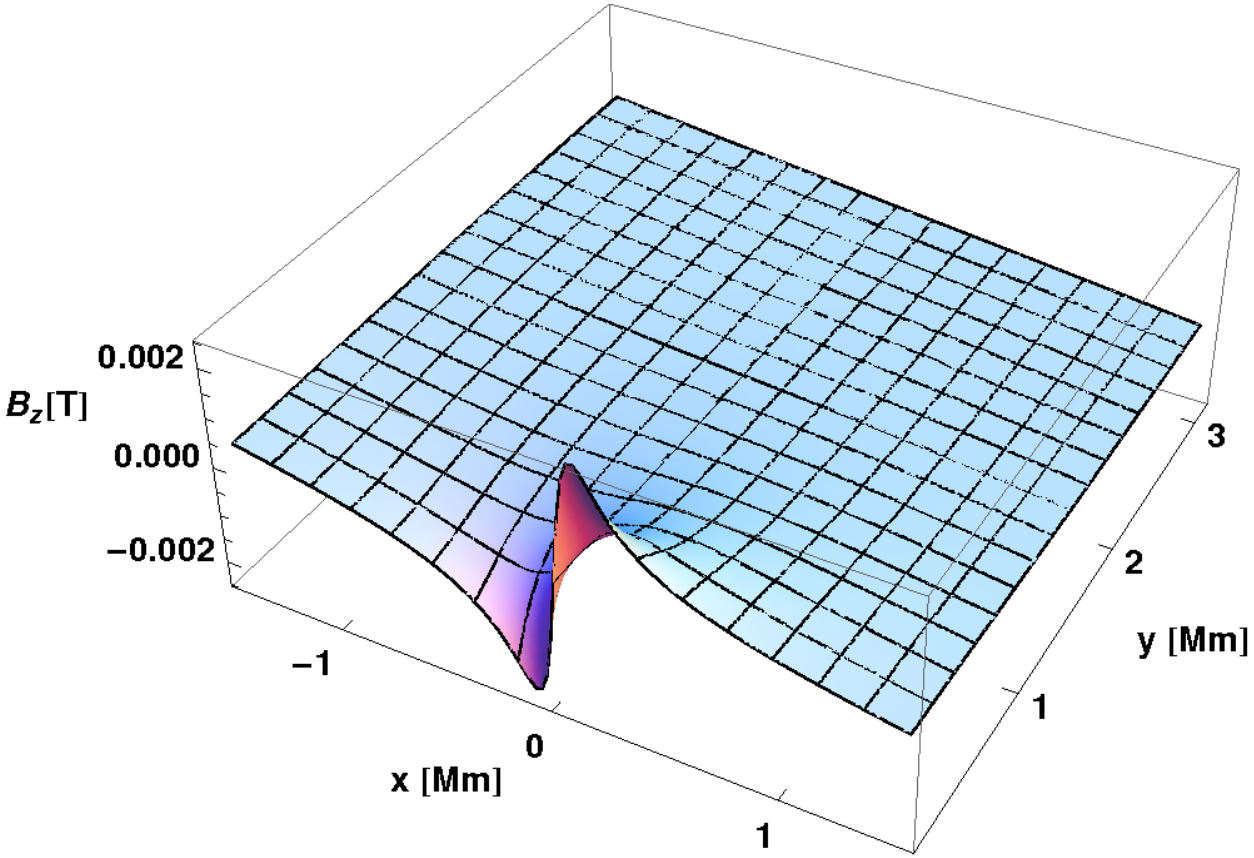}
 \caption{Spatial distribution of the magnetic shear component.}\label{fig:magshear}
 \end{figure}

\begin{figure*}
\centerline{%
\includegraphics[width=0.5\textwidth]{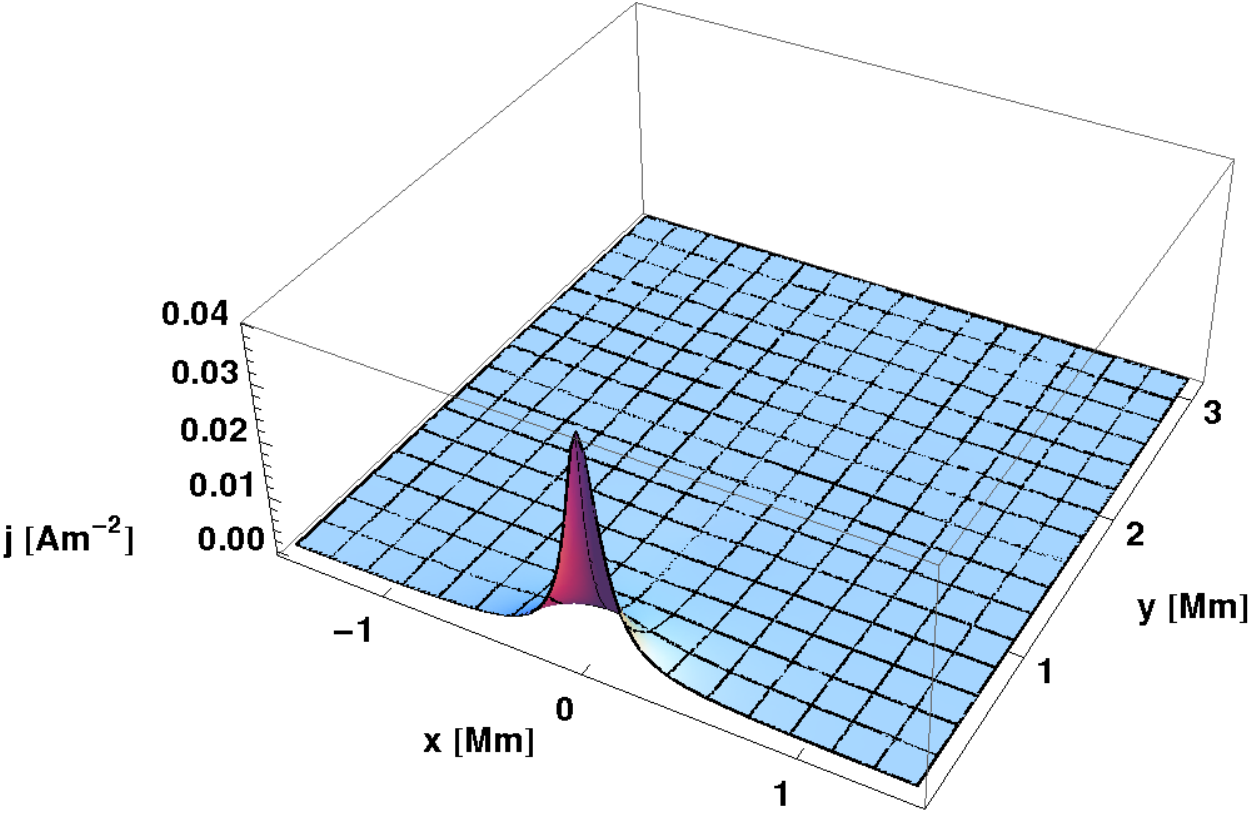}%
\includegraphics[width=0.5\textwidth]{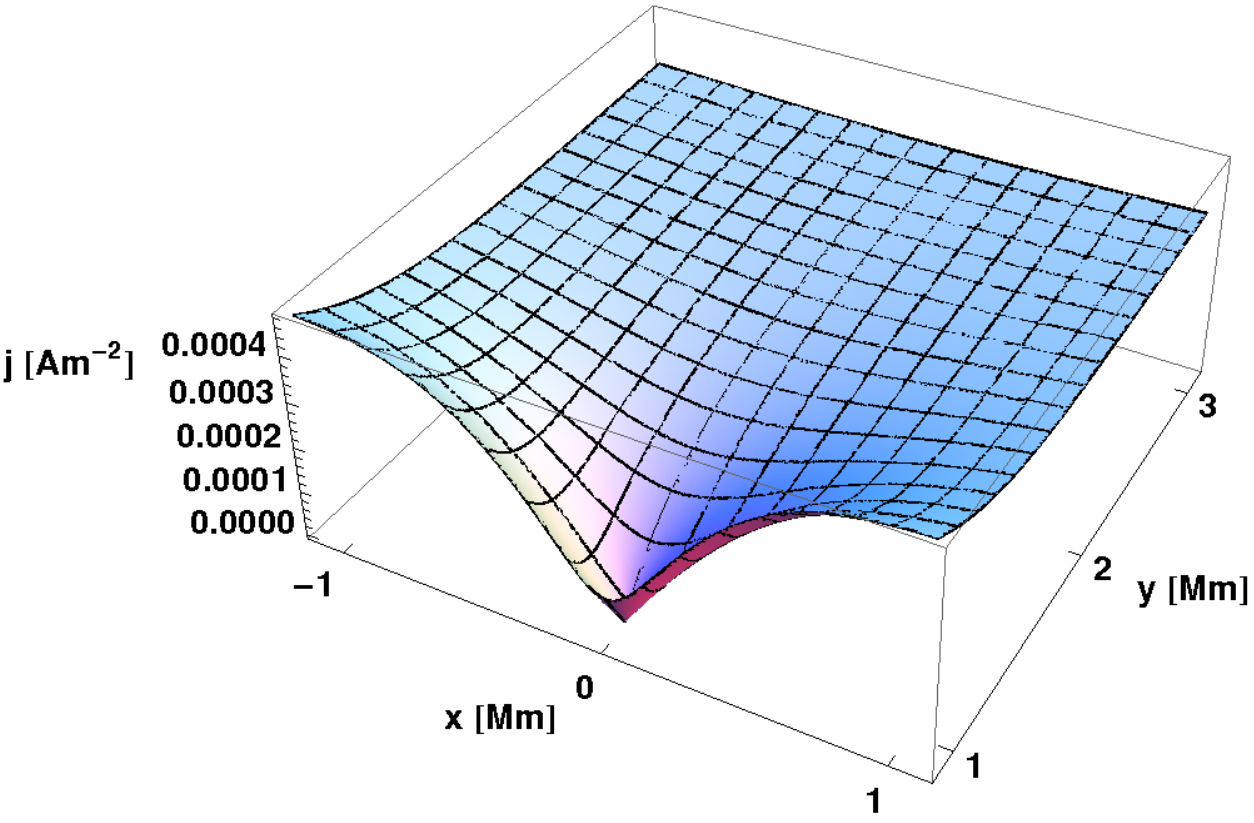}%
}%
\centerline{%
\includegraphics[width=0.5\textwidth]{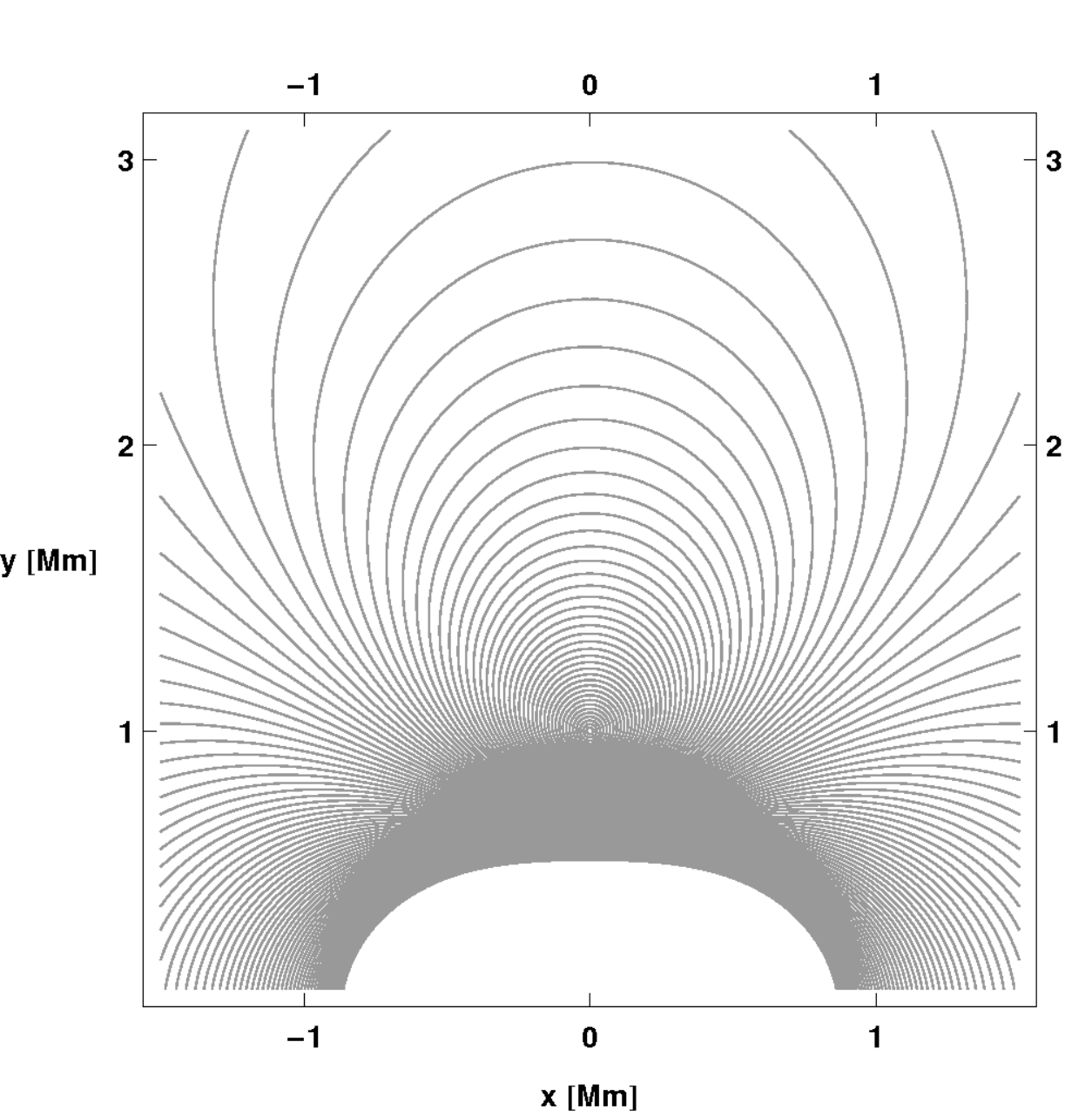}%
\includegraphics[width=0.5\textwidth]{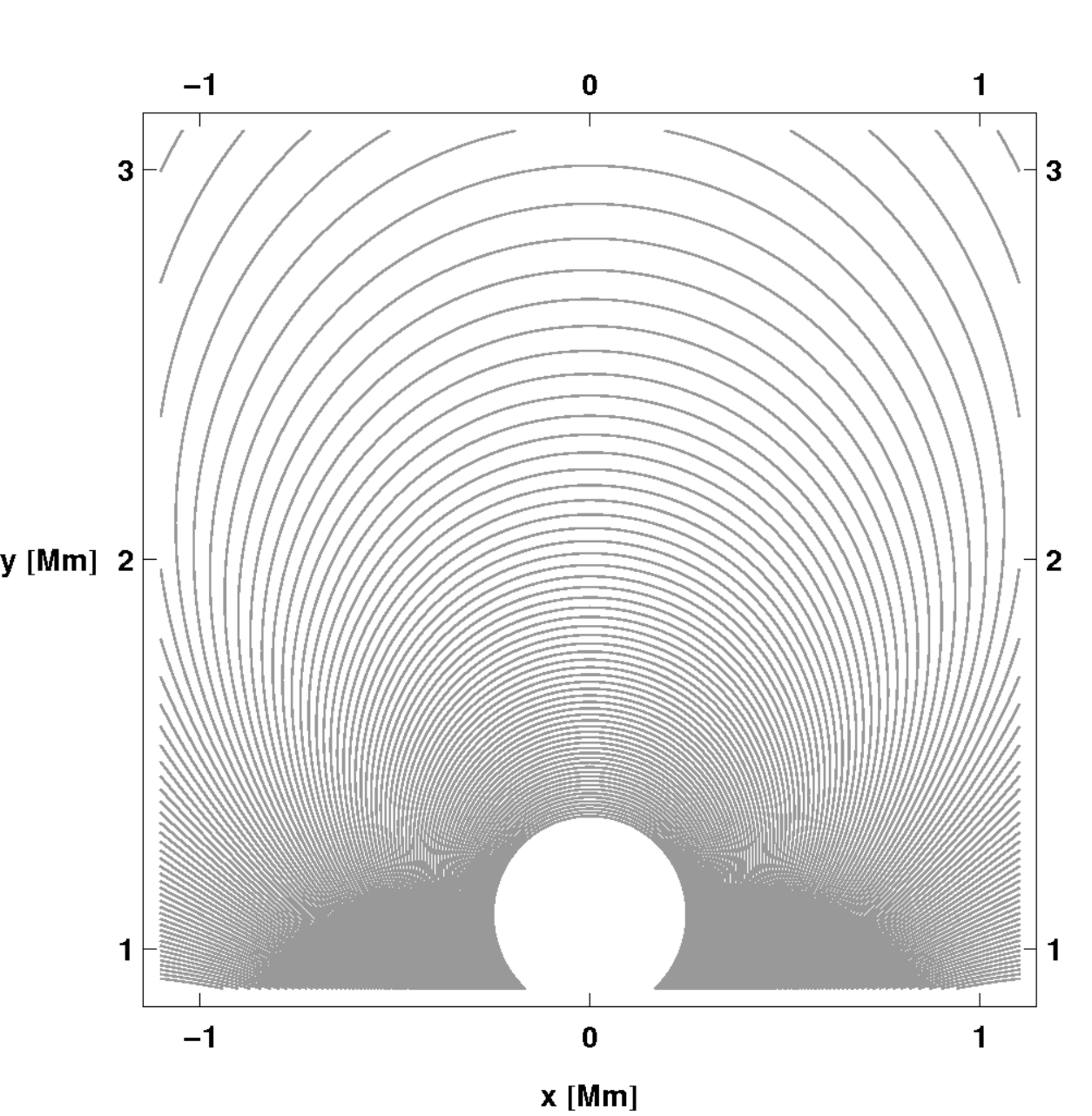}%
}%
\caption{Total current of the stationary state (top) and its contour lines (bottom).
For better visualization the panels to the right display the behavior of the current 
beyond the dipole region, i.e. at far distances.}
\label{fig:current}
\end{figure*}

\begin{figure*}
\centerline{%
\includegraphics[width=0.5\textwidth]{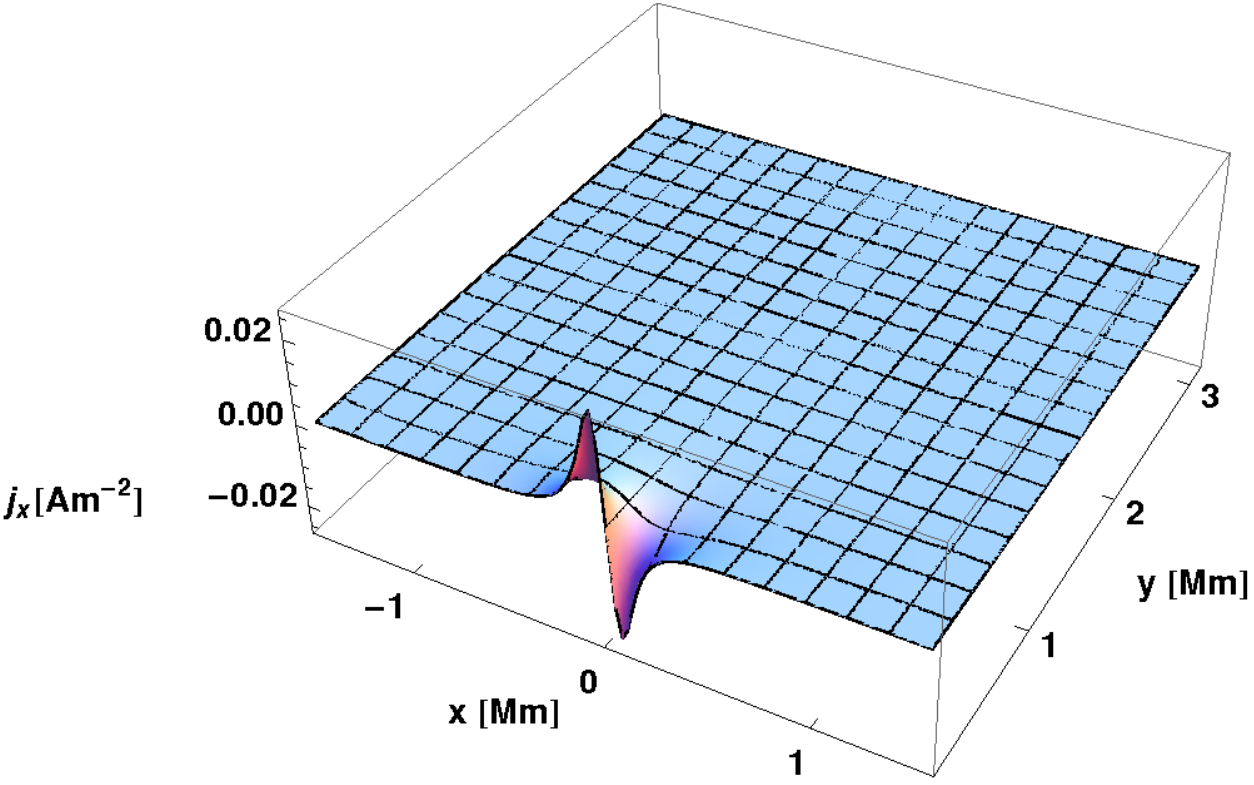}%
\includegraphics[width=0.5\textwidth]{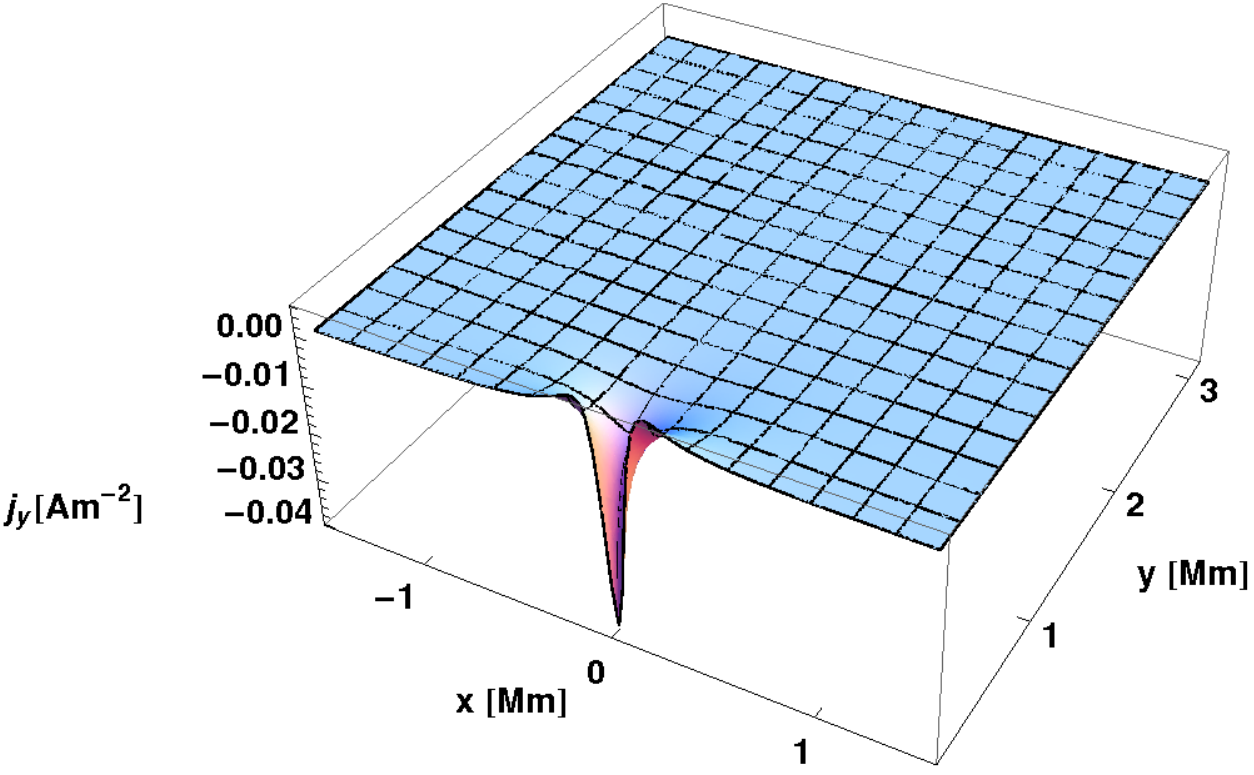}%
}%
\caption{Electric current of the stationary state. Displayed are the $x$-component (left)
and the $y$-component (right).}
\label{fig:currentxy}
\end{figure*}

The crucial term in computing $E_{\parallel}$ is the resistance function $\xi$.
This function was defined in Eq.\,(\ref{elecpot}) to be a pure funtion of $\phi_{m}$. 
However, as was shown in Eq.\,(\ref{xi_phys}), it contains the derivative of the magnetic 
shear component, which is a pure function of $A$. Hence, we need to find a reasonable 
approach for $B_{z}'(A)$,  that allows us to determine the profile of the function 
$\xi(\phi_{m})$. 
According to Eq.\,(\ref{xi_phys}), $\xi$ is given by 
\begin{equation}
\xi(\phi_{m}) = \varepsilon \frac{|\vec B|}{n_{e} e} 
B_{z}'(A)\, . \label{resistance}
\end{equation}
Because we request that the null point of the poloidal magnetic field is also a null point of 
the complete magnetic field, $B_{z}(A=0)$ has to vanish. A Taylor expansion of 
$B_{z}(A)$ around the null point hence delivers that in the lowest order of $A$ the
magnetic shear component must have the form
\begin{equation}
B_{z}(A) = \textrm{const}\cdot A\, ,
\label{const}
\end{equation}
because the contour line $A=0$ is the separatrix. The shape of the magnetic shear component $B_{z}(A)$ is 
shown in Fig.\,\ref{fig:magshear}. For the computation we chose a value for the constant 
in Eq.\,(\ref{const}) of $5\times 10^{-8}$\,m$^{-1}$ to guarantee that $|B_z| < |B_{p}|$. Given the 
highest values of $\sim 10^{-3}$\,T in the dipole region, the magnetic shear component is 
still very small compared with the superimposed poloidal field ($B_{\infty} \sim 10^{-2}$\,T).

The absolute value of the electric current over the considered domain and the 
corresponding contour plot are shown in Fig.\,\ref{fig:current}, while the  
individual contributions in $x$ 
and $y$ direction are shown in Fig.\,\ref{fig:currentxy}. The current flows along the
poloidal field direction, that is, in $x$- and $y$-direction. Above the dipole region, 
here for $y> 1$\,Mm, the total current
is very low and approaches a constant value. This is visible in the top right panel of 
Fig.\,\ref{fig:current}, which shows the total current at large distances, 
and also from the contour plots, which show the increasing separation of the contour lines with
increasing distance from the dipole. 
In the region of 
the dipole field, the $x$-component of the current flows into the positive $x$-direction 
in the left part of the dome, and oppositely in the right part. The dominant component of the
current is the $y$ component, which diverges close to the pole.

The region around the null point can be \lq evacuated\rq, for example, after an ejection of
a flux rope. This means that the total magnetic field and the electron density both approach zero.
Outside the null point region, where the magnetic field saturates, we assume that the density 
saturates as well. Therefore, the term $|\vec B|/(n_{e} e) = |\vec B_{\infty}|/ (n_{e, \infty} e)$ 
can be assumed to be approximately constant. The chosen values for $B_{\infty}$ and $n_{e, \infty}$
can be adjusted to typical coronal values. We emphasize, however, that not every 
arbitrary combination of $|\vec B|$ and $n_{e}$ will successfully deliver a strong enough resistivity
or electric field.
The last unknown in the function $\xi$, which we define 
as the profile function $\xi_{0}(\phi_{m})$, is the term $\varepsilon$. 
This term is requested to cover the $\phi_{m}$ dependence of the resistance $\xi$. For our test 
calculations we set $|\xi_{0}(\phi_{m})| \leq  1$ to guarantee the quasi-neutrality condition and 
the physical significance, in other words, to keep the $\varepsilon$-term bounded as physical
requirement.

For different choices of $B_{z}'(A)$ and $\vec B_{p}$,
we must also recognize that the magnetic shear has to be chosen such that
$|\vec B|\, B'_{z}(A)/(ne)$ results in a reasonable value for the amplitude of the energy per
charge unit, the voltage $\phi_{e}(\phi_{m})$. Otherwise the $\varepsilon$--term has to be adjusted, which
changes the constraint $|\xi_{0}(\phi_{m})| \leq  1$. The implication of a lower value of
$B_{p}$ and/or a higher value of the density has then to be compensated for either by an enhancement of
$B_{z}'(A)$ or an increase of $\varepsilon$, which causes enhanced deviation
from neutrality relative to the deviation from the field-aligned acceleration. This would not cause
problems, because the deviation from neutrality will usually be extremly small, such that any
relative change or fluctuation of the $\varepsilon_{1}$--parameter will be larger. For one single
profile function with $B_{z}'(A)=1/l_{0}$, where $l_{0}$ is a typical lengthscale for the shear that we set to $2\times 10^{7}$\,m, we can write
\begin{equation}
\eta =\varepsilon \frac{l_{0}}{\sigma_{ge}} = 
\xi_{0}(\phi_{m}) l_{0} \left\langle\frac{1}{\sigma_{ge}}\right\rangle
\, .\end{equation}
However, our approach allows defining multiple sites for acceleration and
heating via a fragmented resistivity,
\begin{equation}  
\eta = \sum_{i} \frac{\xi_{0,i}\, l_{0}}{\langle\sigma_{ge}\rangle_{i}} \, .
\end{equation}
The sign of the individual $\xi_{0,i}$ can be either positive or negative,
and accordingly, the direction of acceleration can change at each of these multiple 
acceleration sites.

\begin{figure}
\centering
\includegraphics[width=\hsize]{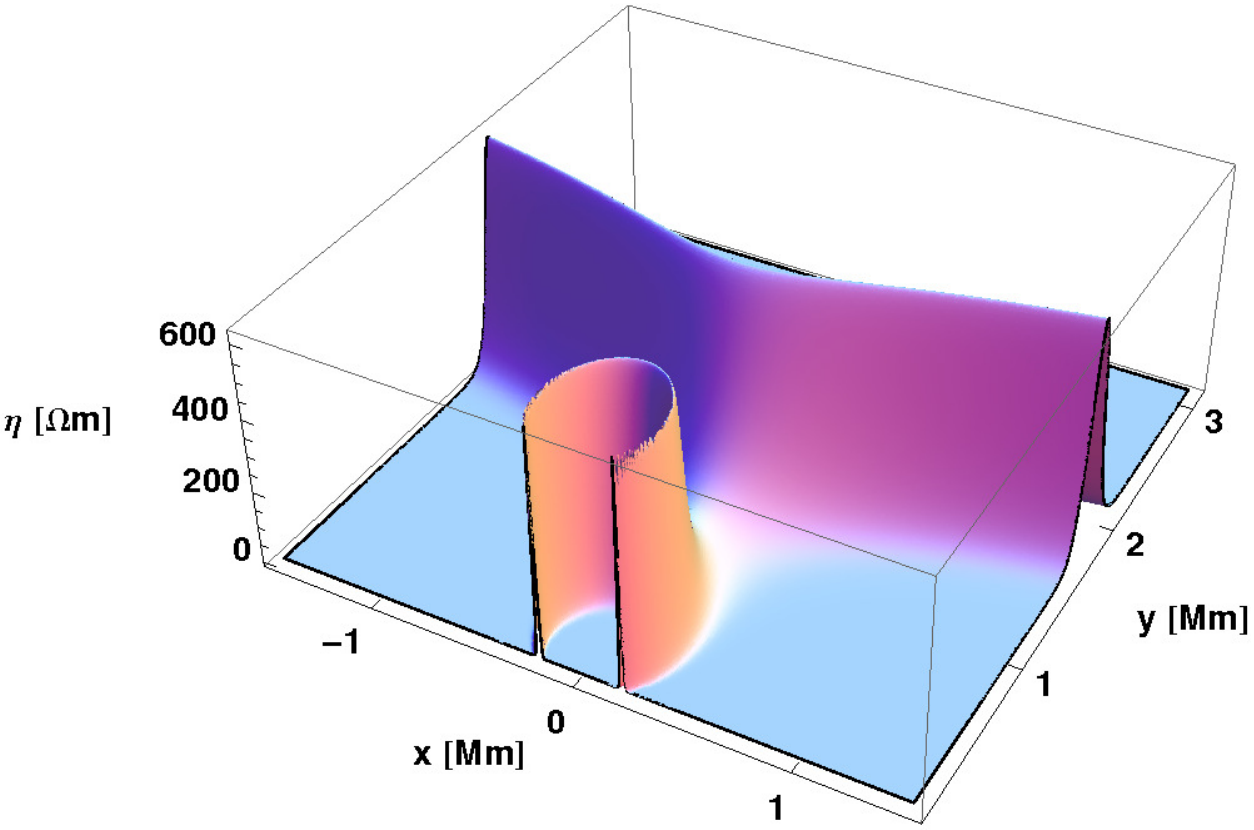}
 \caption{Spatial distribution of the resistivity $\eta$.}\label{fig:resistiv}
 \end{figure}

\begin{figure}
\centering
\includegraphics[width=\hsize]{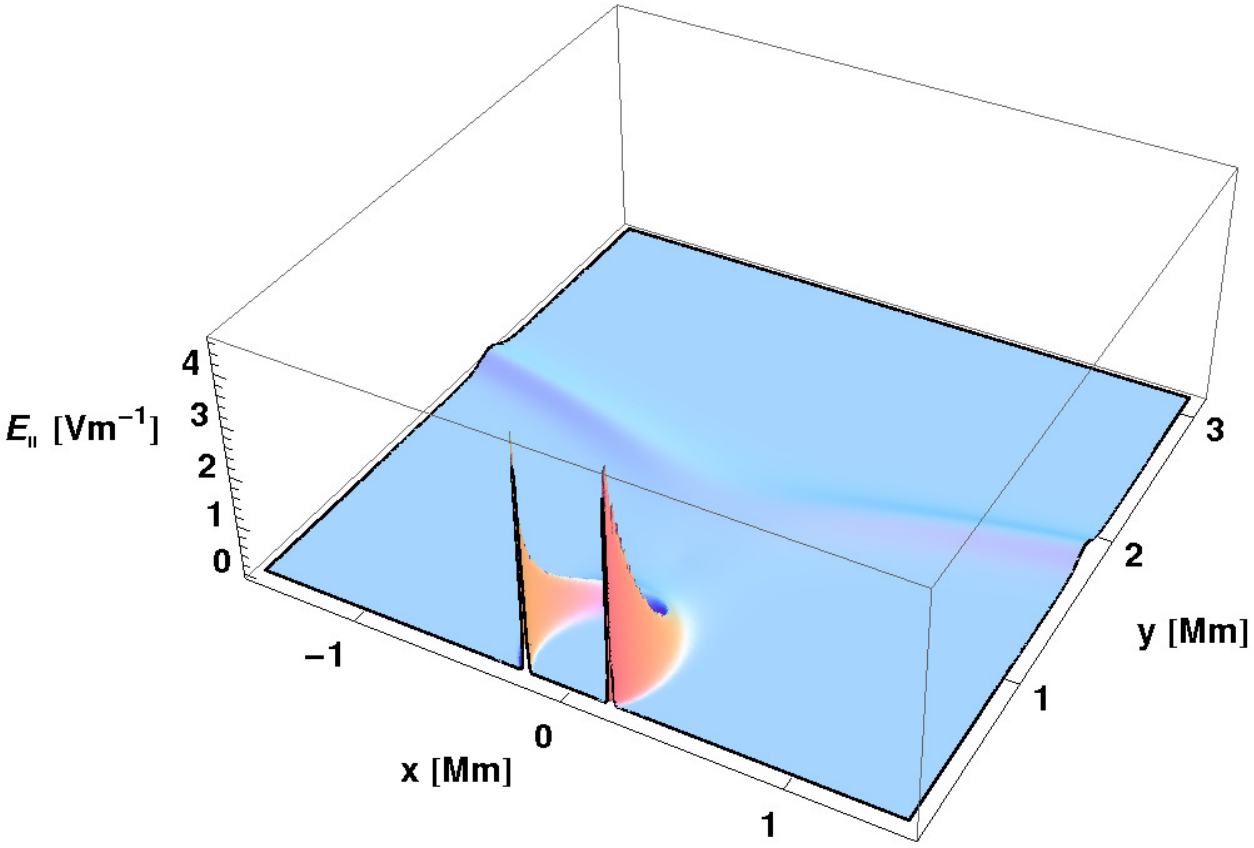}
\includegraphics[width=\hsize]{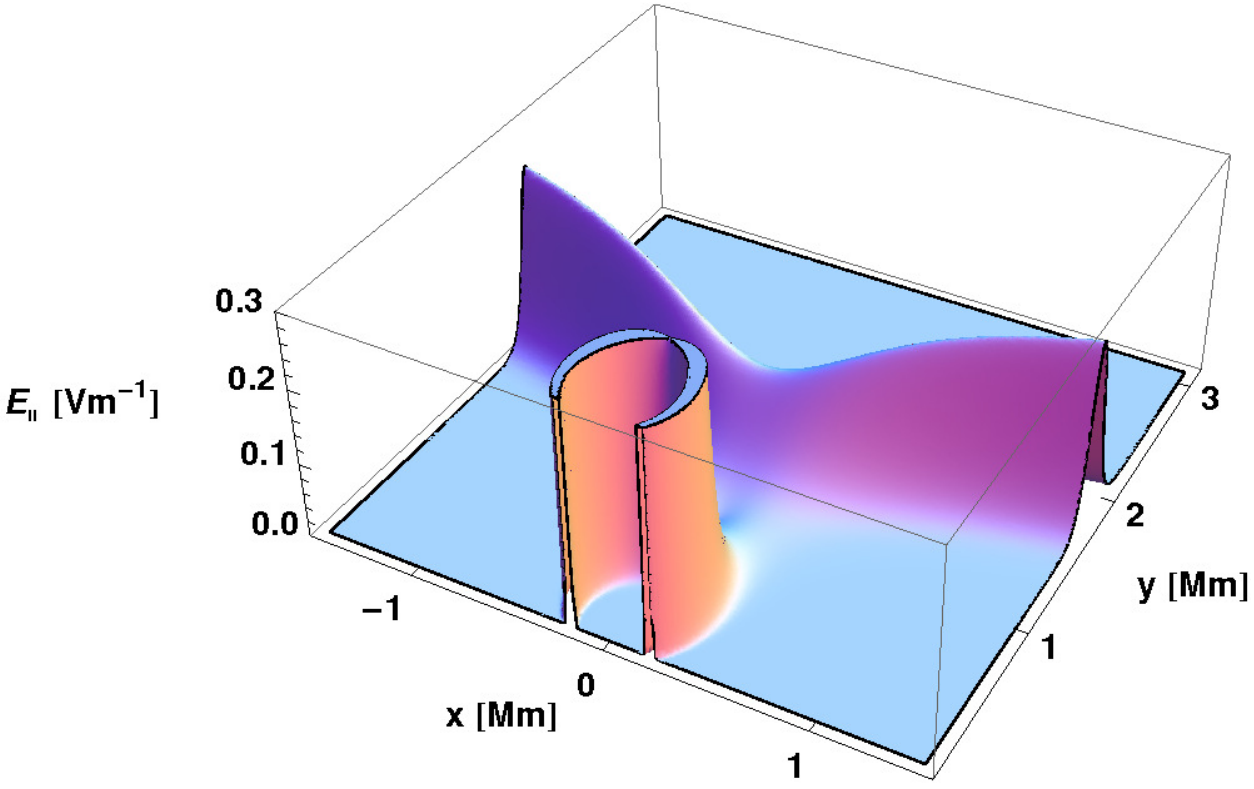}
 \caption{Parallel electric field component. For better visualization, the bottom
panel shows $E_{\parallel}$  cut off at a numerical value of 0.3\,V/m.}\label{fig:epara}
 \end{figure}

In the current investigation, we focused on 
an example using one single,
\lq monolithic\rq~profile of the form
\begin{equation}
\xi_{0} = \cosh^{-2}\left(\frac{\phi_{m}-21500\,\textrm{Tm}}{1000\,\textrm{Tm}}\right)\, .
\end{equation}
From observations \citepads[e.g.,][]{1997ApJ...480..825A}, electron 
densities in acceleration sites of solar flares of $(0.6-10)\times 10^{15}$\,m$^{-3}$ were 
measured. As acceleration regions are usually regions where the 
electron density is preferentially low \citepads[see, e.g.,][]{2002SSRv..101....1A}, 
we fixed it in our model at $n_{e, \infty} = 
10^{14}$\,m$^{-3}$. Fig.\,\ref{fig:resistiv} shows the resulting resisitvity, 
$\eta$, in the computed domain, which for our
chosen simplifications is directly proportional to the total resistance $\xi$. 
The resistivity shows a kinked wall-like structure above the null point region and a 
ring-shaped wall below. This is easily understood from inspecting Eq.\,(\ref{resistance}) and 
the plot of the contour lines of the magnetic potential (Fig.\,\ref{fig:contourphi}). Because the resistivity
is basically the same function as the profile function $\xi_{0}$, which itself is only a function
of the magnetic potential $\phi_{m}$ while all other terms in Eq.\,(\ref{resistance}) are constant,
the resistivity reaches a maximum where $\xi_{0}$ has a maximum. As the isocontours of the function 
$\phi_{m}$ have two disjoint branches at the numerical value of $\phi_{m} = 21\,500$ Tm, every global 
function of $\phi_{m}$ also has two disjoint isocountours with the same isocontour value.

\begin{figure}
 \centering
\includegraphics[width=\hsize]{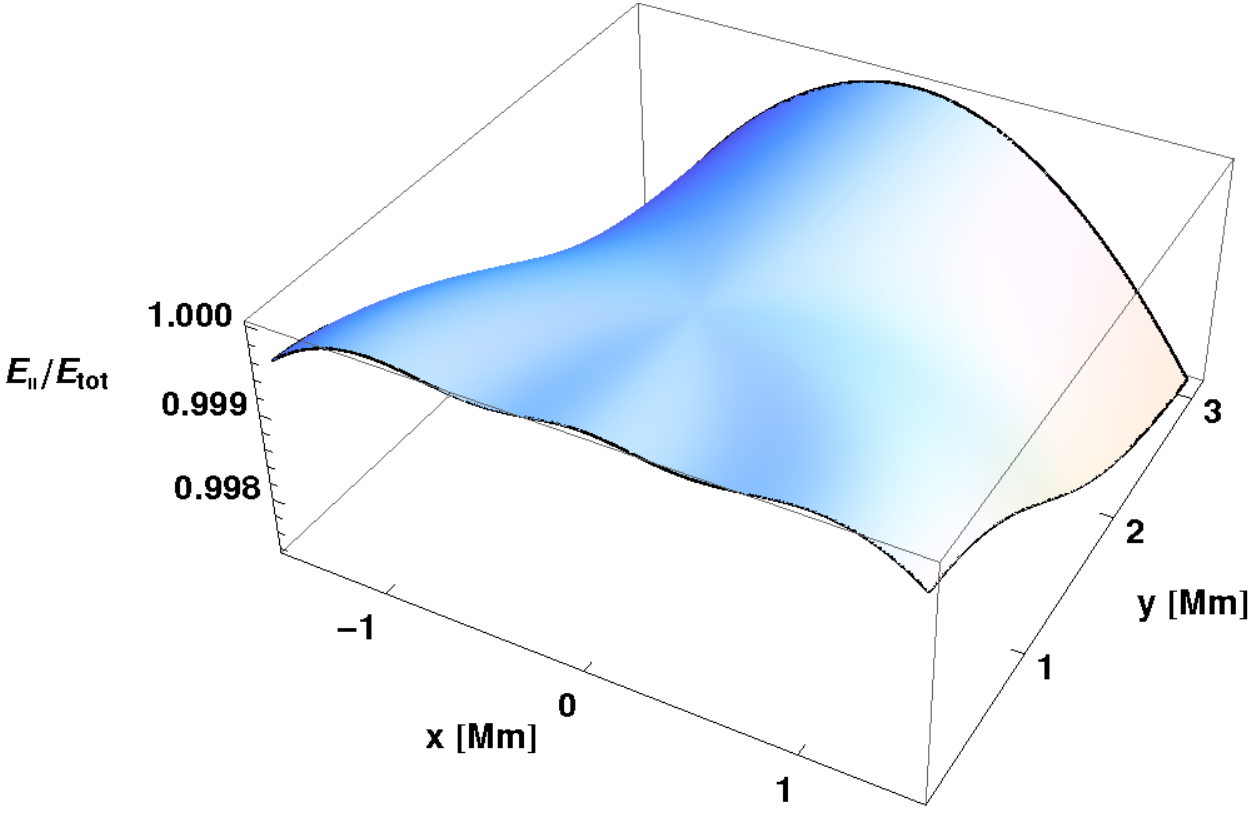}
 \caption{Ratio of the parallel and total electric fields.}\label{fig:elfeldratio}
 \end{figure}

\begin{figure}
 \centering
\includegraphics[width=\hsize]{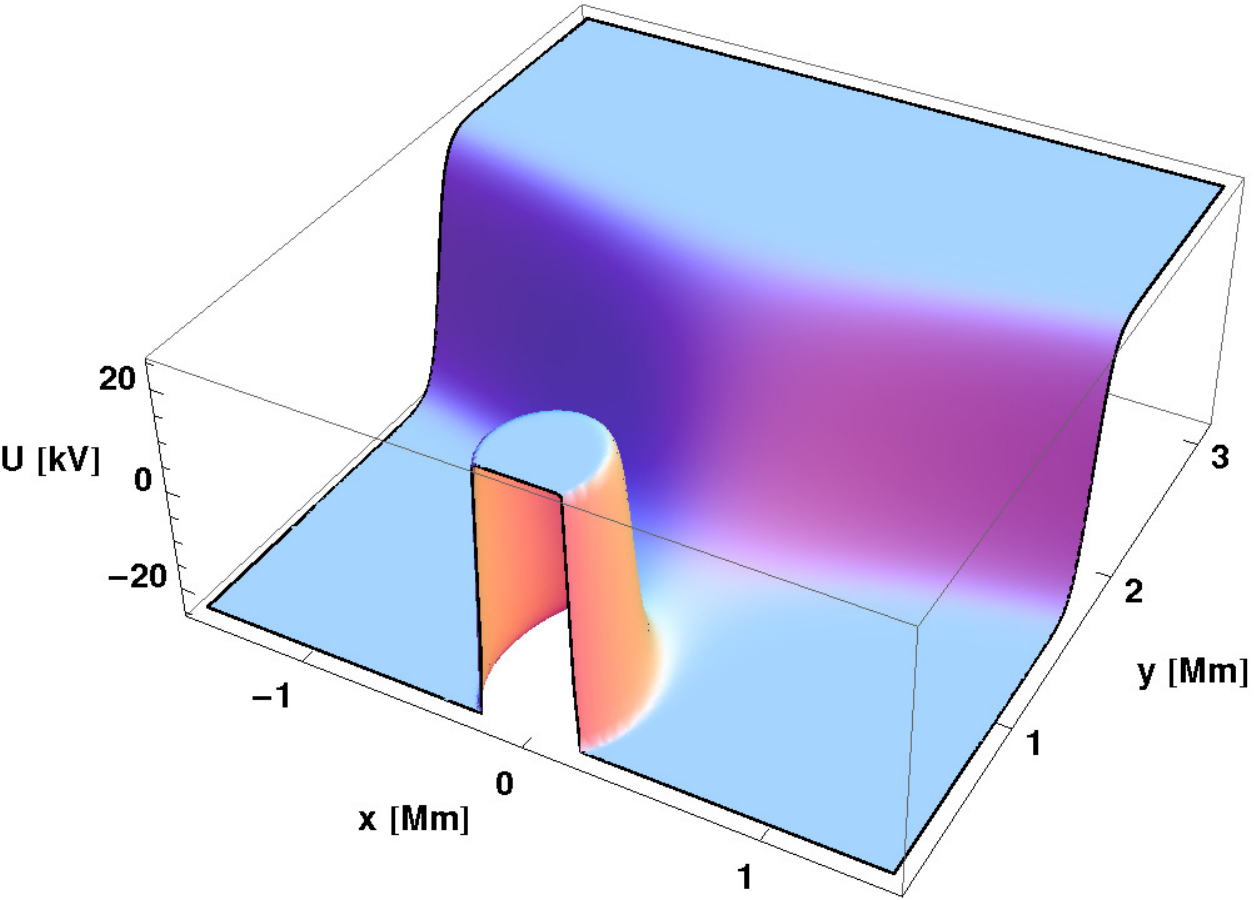}
 \caption{Spatial distribution of the voltage.}\label{fig:voltage}
 \end{figure}

The resulting spatial variation of $E_{\parallel}$ is shown in Fig.\,\ref{fig:epara}.
This parallel component of the electric field is almost identical to the total
electric field (see Fig.\,\ref{fig:elfeldratio}).
Particles are strongest affected, that is, heated and accelerated,
in the domains in which $E_{\parallel}$ is high.
These are also the regions of highest voltage, as is obvious from Fig.\,\ref{fig:voltage}.
Because according to the mathematical frame of our theory 
we are able to superimpose different $\xi_{0}$ profiles, it is possible to construct fragmented 
structures of multiple walls, which provide many regions of enhanced electric field that are suitable for 
consecutive heating (and acceleration) of the particles.

\section{Discussion and conclusions}

A parallel electric field component tends to accelerate particles, especially electrons, out of the thermal 
distribution, resulting in so-called runaway particles. However, the acceleration only works efficiently if the 
parallel electric field component is sufficiently high. Otherwise, the energy gained from the electric field will
mainly be dissipated into heat by dynamical friction (in a collisionless
plasma as in our case) or collisions (in a collisional plasma). Such dissipation connected with Ohmic heating
has been shown to be a reliable mechanism for coronal heating based on
detailed 3D numerical MHD simulations \citepads[see, e.g.,][]{2011A&A...530A.112B, 2013A&A...555A.123B}.

To measure the effectivity of particle acceleration, the parallel electric
field strength $E_{\parallel}$ is typically compared with the Dreicer electric field $E_{\rm D}$
\citepads{1960PhRv..117..329D}. This Dreicer field is clearly
defined in collisional plasmas, where it is on the order of
$E_{\rm D} \simeq 6\times 10^{-4}(n_{14} T_{6}^{-1})$\,V\,m$^{-1}$, where $n_{14}$ is the plasma density in units of
$10^{14}$\,m$^{-3}$, and $T_{6}$ is the temperature in units of $10^{6}$\,K. For collisionless plasmas, the
definition of the Dreicer field is less straightforward and has only been considered for the anomalous resistivity,
where the effective collision frequency is defined, and was found to be 4 to 6 orders of magnitude higher than the classical Spitzer value \citepads[e.g.,][]{1977RvGSP..15..113P, 
2000mare.book.....P}.

In our collisionless scenario with Speiser-like (particle inertia) resistivity, a proper definition 
of a corresponding Dreicer field fails. Because of the nonlinearity between the electric field and the current,
no effective collision frequency can be computed. Instead,
there exists a nonlinear interplay between the particle movement and their relative drift and the geometrical 
structure of the electromagnetic field. Therefore the resistivity does not depend linearly on a collision
frequency (as is the case for the Spitzer and the anomalous or turbulent resistivity),
so that the effective Dreicer field $E_{\rm D_{*}}$ cannot simply be calculated via the relation 
$E_{\rm D_{*}} = (\eta_{*}/\eta_{0})E_{\rm D}$ given by \citetads{1978A&A....68..145N}: In the classical
view, the current density $j= E_{\rm D}/\eta_{0}$ is fixed, so that any enhancement of $\eta_{0}$ causes an
increase of the Dreicer field. 
The inertia-based resistivity approach does not allow fixing the current density and simultaneously enlarging 
the resistivity without enlarging magnetic field and/or reducing parameterically the density.
This is caused by the definition of the differential electric potential $\xi$.
The relation obtained by \citetads{1978A&A....68..145N} assumes that the change of the resistivity
has no influence on the current density. This is not the case in our scenario where, in general, 
the resistivity and the current density depend parametrically on the differential magnetic shear.
Furthermore, Eq.\,(\ref{etadef}) implies that raising the current by raising the differential
magnetic shear $B'_{z}(A)$ alone, the noncollisional resistivity even decreases. 
The assumption that the current is (approximately) fixed implies that $j\propto B_{z}' B_{p}=$ constant
\footnote{As $B_{p}$ is a function of $\phi_{m}$ and $A$ in general, it is not possible to keep
the current fixed.}. Every parametrical increase
of $B'_{z}$ and $B_{z}$ leads to a decrease of $B_{p}$ and therefore also of the electric field,
because $|\vec E| \propto \xi B_{p}$. 
Thus $\xi\propto \sqrt{1+(B_{z}/B_{p})^2} B_{z}' B_{p}$ leads to a slight enhancement
(if $|B_{z}/B_{p}|\stackrel{<}{\sim} 1$) of $\xi$, but the price is the total decrease of
the electric field.
Increasing $B_{p}$ alone would lead to an increase of $\eta$ and of the electric field.

One might doubt the role
of the $\varepsilon_{i}$-terms, of course, but they mainly depend on a geometrical factor, where the
deviation from quasi-neutrality must correspond to the deviation from field-aligned acceleration to avoid the decoupling of $\xi$ from the two-species
system and guarantee the regularity of $\xi$ (bounded value for the electric field and $\xi$). This term can in principle change the Speiser-like resistivity by
order(s) of magnitude, but, as it must be bounded and has to compensate for 
the smallness of the deviation from field-aligned acceleration, it only has a 
marginal influence in our approach.

Hence we have no clear diagnostics at hand to estimate the efficiency of particle acceleration. 
However, according to our initial conditions and requirements (quasi-neutrality, low drift velocity), no strong
acceleration of the bulk particles is expected. Instead, only particles in the high-energy tail of the Maxwellian
distribution might be affected because for them even a (very) small fraction of the classical Dreicer electric
field is sufficient to accelerate them into runaway particles. However, as the resulting electric
field in our model is almost completely parallel to the magnetic field, the particles will experience
some acceleration along the field lines. Our whole scenario is based on a slight charge separation
and a separate treatment of ions and electrons. An ultimate 
investigation of these processes requires a proper two-fluid analysis.

Although a full two-fluid analysis is beyond the scope of the current investigation, we can use the
two-fluid perspective and the parameters from our model calculations to compute
averaged velocities of the particles in the straight field-line 
region\footnote{the region, where the asymptotic boundary condition is reached} 
and estimate from this the highest and lowest energy of the bulk particles in our plasma
model. The electric current density in the straight field-line region is approximately $3\times
10^{-4}$\,A\,m$^{-2}$ (see Fig.\,\ref{fig:current}). On the other hand,
the current density in the two-fluid picture is connected to the
particle velocities via $j = n_{i}qv_{i} - n_{e}ev_{e}$. We assume
that the ion velocity should be on the order of the bulk velocity, which
is $v_{i}\approx  M_{A} B/\sqrt{\mu_{0}\rho}$, where $B\approx 10^{-2}$\, T, 
and $\rho\approx n_{i} m_p$.
Furthermore, the absolute values of the charges of the electron and ion (i.e., in our case protons) are equal,
and the electron and ion densities are approximately
equal because of quasi-neutrality and have (in our model) a value of
$n_{i} \approx n_{e} = 10^{14}$\,m$^{-3}$. This results in an electron 
velocity of
$v_{e}\approx 1/3\,M_{A}\cdot 10^{8}$\, m\,s$^{-1}$. If we assume a minimum Alfv\'en Mach number of 0.1
and a maximum of $\la 1$, the energy of the bulk electrons is between about $1$keV and $10$keV.
In contrast, typical coronal values of the thermal energy of electrons for 
temperatures in the range $10^{6}$K to $10^{7}$K are about $0.1$keV or $1$keV.

Based on our sheared potential field model, we achieve voltage values ($\sim 10$\,kV) 
in agreement with observed
X-ray emission from solar flares \citepads[e.g., with RHESSI, see][]{2002SSRv..101....1A,
2009CEAB...33..141O}. This voltage value can be up to an order of magnitude higher if
we allow for a higher value of $B_{z}'(A)$.
However, the highest voltage is not high enough to produce the highly energetic particles 
with energies in MeV and even GeV range \citepads[see, e.g.,][]{2003ApJ...595L..77H}.
Our calculated example basically produces a 
single wall or sheet (see Fig.\,\ref{fig:epara}), 
meaning that particles are practically accelerated only once. 
For a sustained acceleration of the bulk,
multiple walls are necessary. In the frame of our anlysis, a high number of consecutive walls, even with 
different amplitudes, can be obtained if we allow that either $B_{z}(A)$ or the differential electric potential
$\xi$ or both are functions with a high spatial variation (fragmented). 
Under such considerations the voltage will also be fragmented, producing numerous solitary-wave-shaped \lq walls\rq. 
The existence of such multiple fractal structures within the electromagnetic field allows repetetive acceleration (or decelaration) processes to very high 
energies as well. However, for a spatially variable magnetic shear
component the $\phi_{m}$ dependence of the resistance $\xi$ cannot be expressed in a simple way, and the 
null point of the initial poloidal potential field is not necessarily conserved anymore. The pure linear
dependence of $B_{z}(A)$ on $A$ considered in the presented example is a severe restriction. 
Better models in which $B_{z}(A)$ can be adjusted to constraints and boundary conditions require that in general $\sigma_{g, e}$ must also explicitly
depend on $A$ and $\phi_{m}$, which complicates the analysis. Furthermore,
for a more consistent investigation the generalized Ohm's law needs to be 
considered, and for this the MHD should 
be replaced by a real two-fluid approach.

Nevertheless, the great advantage of our MHD model is that it explicitly 
identifies the current caused by the drift between the accelerated particles 
with the current caused by the magnetic shear component. It is thus a valuable 
and self-consistent approach in the frame of nonideal MHD, which automatically 
incorporates the nonlinear electromagnetic 
feedback of the particles, which is ignored in the usual test particle approach.

%


\begin{acknowledgements}
We thank the anonymous referee for useful comments and suggestions on the draft.
This research made use of the NASA Astrophysics Data System (ADS). D.H.N. 
and M.K. acknowledge financial support from GA\,\v{C}R under grant numbers 13-24782
and P209/12/0103, respectively. The Astronomical Institute Ond\v{r}ejov is 
supported by the project RVO:67985815.
\end{acknowledgements}

%
%
\bibliographystyle{aa}
\bibliography{efield_v2}  
\end{document}